\documentclass[aip,jcp,amsmath,reprint,notitlepage,superscriptaddress,a4paper,floatfix]{revtex4-1}
\usepackage{graphicx}
\usepackage{amssymb}
\usepackage{amsmath}
\usepackage{amsfonts}
\usepackage{color}
\usepackage{bm}
\usepackage[update,prepend]{epstopdf}

\newcommand{\la}{\left\langle}
\newcommand{\ra}{\right\rangle}
\begin{document}
\title{Molecular simulations of phase separation in elastic polymer networks}

\author{Takahiro Yokoyama}
\affiliation{Leibniz-Institut f{\"u}r Polymerforschung Dresden e.V., Hohe Stra\ss e 6, 01069 Dresden, Germany}
\affiliation{Institut f{\"u}r Theoretische Physik, Technische Universit{\"a}t Dresden, 01069 Dresden, Germany}

\author{Yicheng Qiang}
\affiliation{Max Planck Institute for Dynamics and Self-Organization, Am Fa\ss berg 17, 37077 G{\"o}ttingen, Germany}

\author{David Zwicker}
\email{david.zwicker@ds.mpg.de}
\affiliation{Max Planck Institute for Dynamics and Self-Organization, Am Fa\ss berg 17, 37077 G{\"o}ttingen, Germany}

\author{Arash Nikoubashman}
\email{anikouba@ipfdd.de}
\affiliation{Leibniz-Institut f{\"u}r Polymerforschung Dresden e.V., Hohe Stra\ss e 6, 01069 Dresden, Germany}
\affiliation{Institut f{\"u}r Theoretische Physik, Technische Universit{\"a}t Dresden, 01069 Dresden, Germany}
\affiliation{Cluster of Excellence Physics of Life, Technische Universit{\"a}t Dresden, 01062 Dresden, Germany}

\date{\today}

\begin{abstract}
Phase separation within polymer networks plays a central role in shaping the structure and mechanics of both synthetic materials and living cells, including the formation of biomolecular condensates within cytoskeletal networks. Previous experiments and theoretical studies indicate that network elasticity can regulate demixing and stabilize finite-sized domains, yet the microscopic origin of this size selection remains elusive. Here, we use coarse-grained molecular dynamics simulations with implicit solvent to investigate how network architecture controls phase separation and limits domain growth. By systematically varying chain contour length, chain rigidity, and network topology, we uncover that finite domains emerge when intrinsic chain- or network-level length scales, such as persistence length or entanglement length, impose local constraints on coarsening. Further, the size of these finite domains is highly correlated with these microscopic network properties, but depends surprisingly little on bulk elasticity. Taken together, our findings establish a molecular basis for understanding droplet formation in polymer networks, and provide guiding principles for engineering materials and interpreting condensate behavior in cells.
\end{abstract}

\maketitle
\section{Introduction}
Phase separation in polymeric materials occurs in diverse settings, including solutions,\cite{wang:advmat:2019, yuan:acsnano:2023} melts,\cite{feng:po:2017, bates:ma:2017, matsen:jcp:2020} and cross-linked gels,\cite{wang:pps:2024, yang:chem:2023, style:prx:2018, kim:sciadv:2020, sicher:sm:2021, fernandez:jacsau:2022} where unfavorable interactions between components drive demixing into domains enriched in different species.
Typically, such systems undergo \textit{macrophase separation}, as in oil-water mixtures, where domains coarsen indefinitely to minimize the interfacial energy between the two immiscible liquids. 
Finite-sized domains only emerge when additional constraints are imposed, leading to \textit{microphase separation} with stable structures such as lamellae, droplets or bicontinuous morphologies, which is well established in block copolymers and surfactant assemblies.\cite{ghosh:biochem:2020, lee:angew:2023, panagiotopoulos:ma:2024}
In those systems, domain sizes are tightly coupled to the architecture of the nanoscale building blocks, e.g., the block length in copolymers, placing inherent limits on the achievable domain sizes. 
In contrast, polymer networks potentially provide greater freedom, as their (local) elasticity, topology and entanglements introduce emergent length scales beyond those of individual chains.
In synthetic polymer gels, for example, these mechanisms enabled patterned microstructures with finite characteristic length scales.\cite{fernandez:nmat:2024}
In living cells, finite-sized biomolecular condensates\cite{brangwynne:sci:2009, hyman:arcdb:2014, banani:nrevcellbio:2017} form through liquid-liquid phase separation of proteins and other biomolecules within active, viscoelastic\cite{tanaka:comphys:2022} and crowded\cite{shu:prxlife:2024} environments, such as cytoskeletal and chromatin networks.
Despite this broad relevance, the mechanisms of phase separation within polymer networks remain incompletely understood, owing to the additional network-imposed constraints and interactions between the embedded liquid and the surrounding scaffold.\cite{wiegand:etls:2020, lee:nphys:2021, lee:bioeng:2022, boddeker:nphys:2022, boddeker:prxlife:2023, mohapatra:brb:2023, liu:ncommun:2023}

Recent experiments\cite{fernandez:nmat:2024} have demonstrated microphase separation in flexible networks, with domain sizes increasing from hundreds of nanometers to micrometers, often exceeding characteristic single-chain dimensions. 
Although domain size grows with decreasing network stiffness, it remains unclear whether elasticity itself is the direct cause, or merely correlates with other microscopic constraints. 
Theory\cite{wei:prl:2020, rosowski:sm:2020, rosowski:nphys:2020,vidal-henriquez:sm:2020, vidal-henriquez:pnas:2021, kothari:jmps:2020, kothari:bmm:2023, biswas:sm:2022, ronceray:epl:2022, meng:pnas:2024} and simulations\cite{zhang:prl:2021, curk:pnas:2023} may help clarify this missing link by identifying the molecular mechanisms and length scales that govern domain growth. 
Classical field theories\cite{cahn:jcp:1958,cahn:am:1961} capture the early-time dynamics of spinodal decomposition and domain coarsening driven by interfacial energy, but cannot explain the arrest of demixing.
Achieving microphase separation requires free energy functionals that include long-range interactions;\cite{ohta:ma:1986, glotzer:pre:1994, muratov:pre:2002, kumar:prl:2023} in block copolymer melts,\cite{ohta:ma:1986} covalent connectivity provides a natural long-range penalty, producing finite domains on the scale of the polymer blocks.
Similar approaches have been applied recently to elastic polymer networks, where introducing a non-local length scale into the elastic free energy reproduces microphase separation.\cite{qiang:prx:2024, mannattil:prl:2025, oudich:jmps:2026}
These studies emphasize that a network-associated length scale is essential for arresting phase separation, though its microscopic physical origin has remained elusive.

To elucidate which microstructural features of the polymer network govern phase separation and how, it is necessary to consider the multiple contributions that determine its mechanical response. 
Elasticity arises from a combination of factors,\cite{rubinstein:ma:2002, gula:ma:2020, sorichetti:ma:2021, tian:ma:2025} including bond and bending rigidity of individual chains, entropic elasticity associated with chain conformations, network connectivity set by cross-linking density, and topological constraints.
These contributions give rise to multiple, potentially overlapping length scales, which can combine in emergent ways to create new characteristic scales not evident from the individual components.
In synthetic polymer gels composed of flexible polymers, elasticity is dominated by entropic contributions, network connectivity, and topological constraints.
In contrast, biopolymer networks such as the cytoskeleton often consist of semi-flexible filaments, e.g., actin filaments and microtubules, where chain rigidity plays an additional important role.\cite{broesersz:revmodphys:2014, burla:nrevphys:2019, schepers:pnas:2021, lorenz:bprev:2022}
Consequently, the pathways through which these networks regulate phase separation might differ, reflecting the distinct length scales as well as the energetic and entropic contributions specific to each system.

These observations motivate a focused investigation of which microscopic length scales arrest phase separation, and how they relate to network elasticity. 
Molecular simulations are ideally suited for this task, because they provide direct control over microscopic chain and network properties. 
Using a particle-based approach, we construct polymer network models with varied architectures and systematically tune chain contour length, chain rigidity, and topological constraints to examine their impacts on macroscopic elasticity and phase separation behaviors.
By comparing different network architectures, we identify the network-specific length scales that arrest phase separation and quantify how they modulate the domain sizes. 
This study bridges molecular architecture and continuum models, clarifies why domain sizes in elastic networks remain finite, and offers design rules for creating synthetic and biological polymeric scaffolds.

\section{Results and Discussion}
\subsection{Preparation of Polymer Networks with Varying Length Scales}
Polymer networks exhibit several intrinsic length scales, reflecting both material-  and network specific features. 
For an individual chain, key scales are the segment length $L_{\text{b}}$, which is the distance between two subsequent monomeric units, the persistence length $L_{\text{p}}$, which describes the bending rigidity, and the contour length $L_\text{c}$.
When these chains are linked into a network, additional length scales emerge, such as the distance $a$ between two adjacent cross-links, and the entanglement length $L_{\text{e}}$, which quantifies the typical distance between successive topological constraints.
These characteristic length scales govern the macroscopic mechanical and rheological properties of the network, yet they are often challenging to identify unambiguously in experiments.

\begin{figure*}[htb]
    \centering
    \includegraphics[width=12cm]{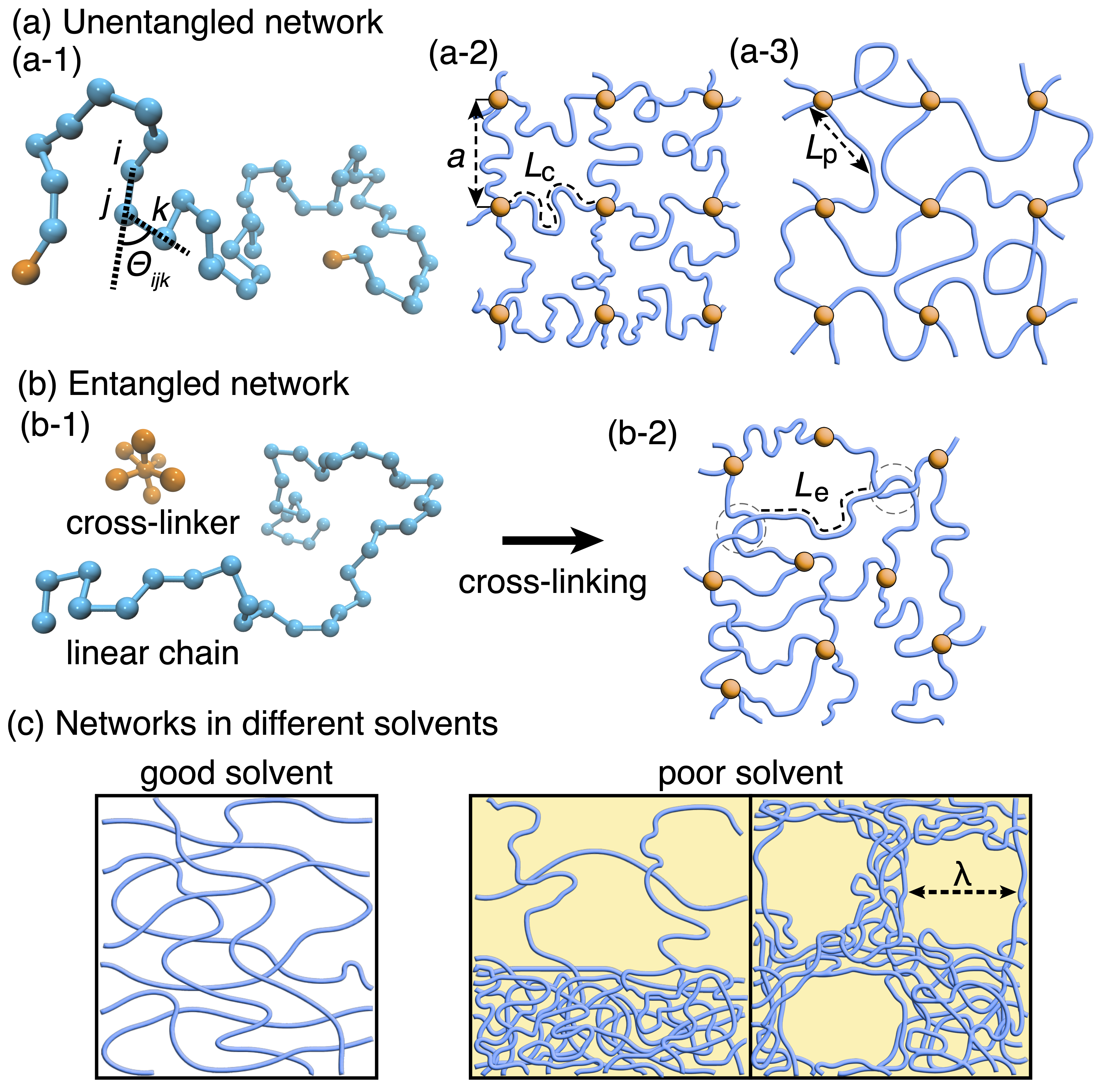}
    \caption{(a) Preparation of regular networks. (a-1) Conformation of an individual network strand, where blue and orange particles represent monomers and cross-linkers, respectively. Two-dimensional schematics of the regular networks under the condition of (a-2) $L_{\text{p}} \ll a \ll L_{\text{c}}$, and (a-3) $a \lesssim L_{\text{p}} < L_{\text{c}}$. (b) Preparation of entangled networks. (b-1) Molecular structures of a hexa-functional cross-linker and a linear chain. (b-2) Two-dimensional schematic of an entangled network after the cross-linking procedure. (c) Schematic of networks in different solvents: homogeneous phase in good solvent (left panel), macro- or microphase separation in poor solvent (right panel).}
    \label{fig:fig1}
\end{figure*}

Our first objective is to produce polymer networks with systematically varied characteristic length scales.
To this end, we employ a particle-based method using a coarse-grained bead-spring model, in which spherical monomers of diameter $\sigma$ are connected by finitely extensible nonlinear elastic (FENE) bonds, enabling control over these different length scales.
In our simulations, the unit of length is set to $\sigma$, which can be mapped to real units as $\sigma \approx 1\,\text{nm}$ when considering the typical segment length of synthetic polymers and intrinsically disordered proteins, or as $\sigma \approx 5-30\,\text{nm}$ when considering supramolecular fibers such as filamentous-actin or microtubuli.
The contour length $L_{\text{c}}$ can be tuned via the cross-linking fraction $\phi_\text{c}$ in the system, where $\phi_\text{c}$ is the number of cross-linked monomers divided by the total number of monomers; generally, $L_\text{c}$ decreases with growing $\phi_{\text{c}}$.
The persistence length $L_{\text{p}}$ is controlled through a harmonic bending potential given by Eq. (\ref{eq:UBEND}), which depends on the angle $\Theta_{ijk}$ defined by three consecutive monomers $i$, $j$, and $k$, and is modulated by the rigidity parameter $\kappa$ (Fig. \ref{fig:fig1}(a-1)).
We determined the persistence length from $L_\text{p}=-L_\text{b}/\ln{\langle \cos{\Theta_{ijk}} \rangle}$, which reflects the exponential decay of bond orientation correlations along the polymer contour.\cite{rubinstein:book:2003}
For sufficiently large values of $\kappa/k_{\text{B}}T \gtrsim 2$, the persistence length of a single, free chain is $L_{\text{p,0}}/L_{\text{b}} \approx \kappa / k_{\text{B}}T$ due to the equipartition theorem, where $k_{\text{B}}$ is Boltzmann's constant and $T$ is the temperature.\cite{nikoubashman:jcp:2016, milchev:jcp:2018, midya:jcp:2019, nikoubashman:jcp:2021}
In contrast, the entanglement length $L_{\text{e}}$ is an emergent property rather than an input parameter, which makes it difficult to control directly.
Thus, we employ two distinct network preparation schemes to generate networks with and without entanglements (see below and Sec. S1 in the SI for details).

In the first scheme (``regular networks''), we construct defect-free networks by placing cross-linking junctions at the vertices of a simple-cubic lattice, and connecting each vertex to its six nearest neighbors with a polymer chain [Fig. \ref{fig:fig1}(a)]. Note that the cross-linking points are not fixed in space but can move as the other monomers. This procedure yields networks with uniform strand metrics and mesh size, enabling systematic investigation of the network-specific length scales.
The lattice constant $a$ is the direct distance between two neighboring cross-links, with strand contour length $L_{\text{c}} > a$ by geometry [Fig. \ref{fig:fig1}(a-2)].
This setup is also well suited for studying the influence of the strand stiffness, since the ratio $L_\text{p}/L_\text{c}$ is uniform within the network [Fig. \ref{fig:fig1}(a-3)].
This construction excludes topological entanglements by design; the only permanent constraints are the cross-linking junctions, so the measured response reflects cross‑link density and chain stiffness rather than entanglement effects.

In the second scheme (``entangled networks''), we use a complementary approach to generate entangled networks, beginning with a mixture of linear polymer chains and free hexa-functional cross-linkers  [Fig. \ref{fig:fig1}(b-1)].
After allowing the system to relax into a homogeneously mixed state, we bind the functional groups of the cross-linkers to the nearest monomers of the linear chains.
This protocol yields a disordered network that preserves the entanglements formed in the initial polymer solution.
Such networks are particularly suited for investigating characteristic length scales that arise from topological constraints within the network, in particular the entanglement length $L_{\text{e}}$ [Fig. \ref{fig:fig1}(b-2)], alongside cross-link-controlled scales.

To probe the effects of the different polymer length scales, we prepared three types of polymer networks: (i) regular networks composed of flexible chains, (ii) regular networks composed of semi-flexible chains, and (iii) entangled networks with flexible chains.
Every network is defined by the strand contour length $L_\text{c}$ and the (average) distance between neighboring cross-linking sites $a$; in addition, case (ii) is characterized by the persistence length $L_\text{p}$, while case (iii) depends on the entanglement length $L_{\text{e}}$.
To facilitate the comparison between the different networks, we used the same monomer number density $\rho = 0.4 \sigma^{-3}$ (corresponding to a monomer volume fraction $\phi \approx 0.21$) in all cases.
Further, we verified that all networks are fully percolated, thus exhibiting macroscopic connectivity through the periodic boundary conditions.\cite{peleg:epl:2007, peleg:ma:2008}
For the regular networks, the cross-linking fraction $\phi_\text{c}$ was controlled by changing the number of monomers $N$ between the cross-linking points in the range of  $10 \leq N \leq 40$, resulting in a strand contour length of $L_{\text{c}}=NL_{\text{b}}$ and lattice constant $a = \left[(3N-2)/\rho\right]^{1/3}$.
The cross-linking fraction is then calculated as $\phi_{c}=6/(3N-2)$, which varies between $0.0508 \leq \phi_{\text{c}} \leq 0.2143$ depending on $N$.
Regular networks with semi-flexible strands were prepared by gradually increasing the bending rigidity parameter $\kappa$ until the desired stiffness was achieved. 
Note that equilibrating these semi-flexible networks is highly challenging due to the mismatched length scales ($a \ll L_{\text{c}}$ and $a \lesssim L_{\text{p}}$), which forced filaments into unfavorable conformations that relax only slowly because of the imposed cross-linking constraints (see Secs. S2-S4 in the SI for details).
Such slow coarsening was also reported in another recent study of fibrillar gels formed by semi-flexible polymers.\cite{kroeger:sm:2025} 
In the entangled networks, the chain length was fixed to $N=40$ and the cross-linking fraction $\phi_\text{c}$ was controlled by adjusting the number of activated functional groups on the cross-linkers.
When all functional sites were connected to monomers, the resulting cross-linking fraction was $\phi_\text{c}=0.1034$, which is identical to that of the regular network with $N=20$.

The solvent is modeled implicitly, and effectively occupies the volume not occupied by the monomers and cross-linkers [Fig. \ref{fig:fig1}(c)].
The solvent quality is incorporated into the monomer-monomer interaction. 
In a good solvent, monomers experience effective repulsion -- modeled using the Weeks-Chandler-Andersen (WCA) potential\cite{weeks:jcp:1971} -- to maximizes their exposure to the fictitious solvent.
Conversely, poor solvent causes effective attraction between monomers -- described by the standard Lennard-Jones (LJ) potential -- to reduce unfavorable interactions with the surrounding implicit solvent.
Good solvent conditions were employed for the initial equilibration, resulting in a swollen polymer network [Fig. \ref{fig:fig1}(c)], from which we determined the characteristic length scales as well as the elastic properties of the networks.
Phase separation was then triggered by gradually lowering the solvent quality, which corresponds to cooling down for systems with an upper critical solution temperature.
We monitored the potential energy and the conformations of the constituent polymers, and took measurements only after these quantities reached a steady plateau.
The phase separation length scale was determined by the size of the network-dilute regions, and the microphase separation length scale $\lambda$ was defined only for systems with arrested domain coarsening [Fig. \ref{fig:fig1}(c)]. Additional technical details are provided in Sec.~\ref{sec:method} and the SI.

\subsection{Phase Separation Modulated by Network Architecture}
\label{subsec:phase_separation}
\begin{figure*}[htb]
    \centering
    \includegraphics[width=13cm]{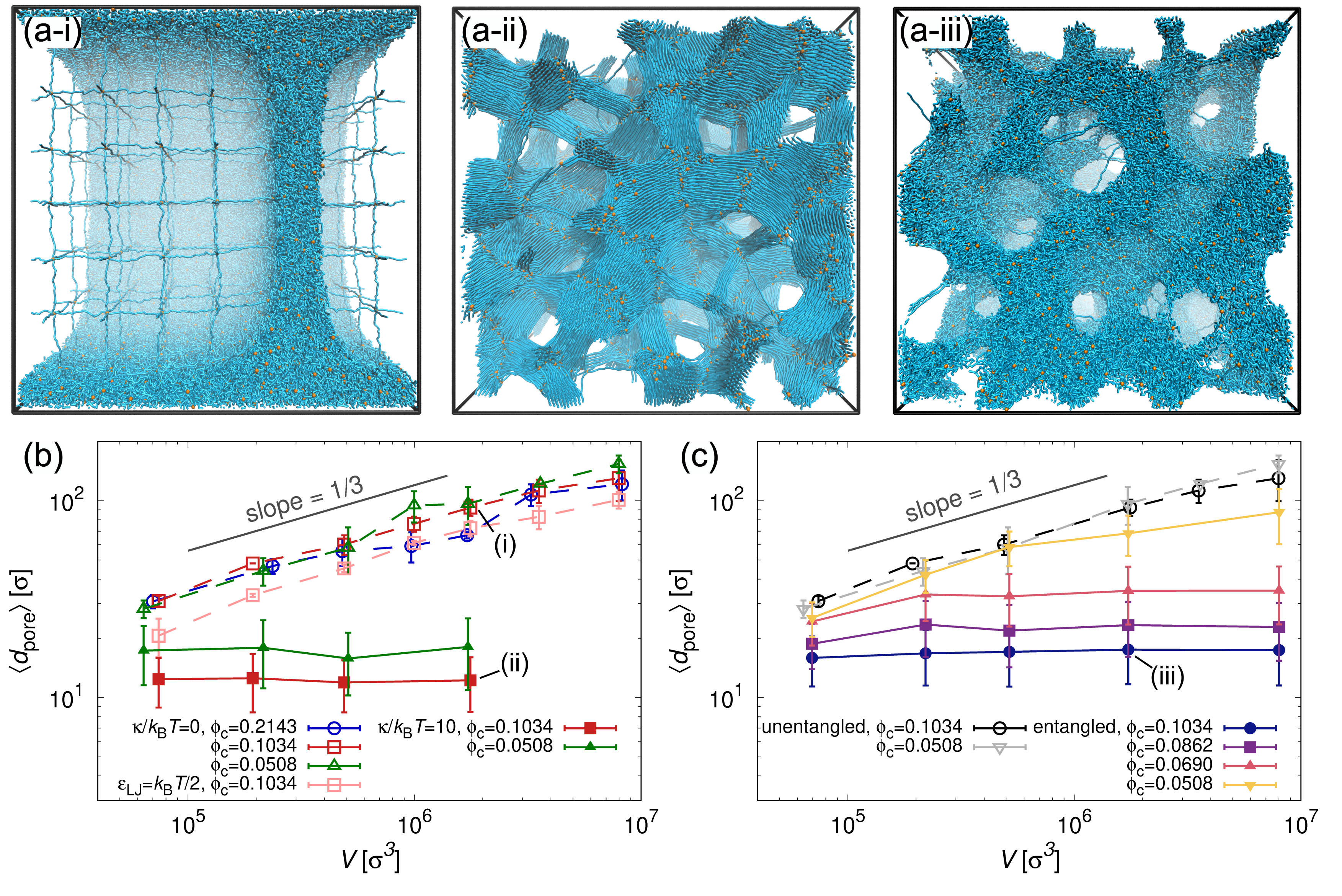}
    \caption{(a) Representative snapshots of three different phase separation behaviors for (i) regular network with flexible chains ($\kappa/k_{\text{B}}T=0$), (ii) regular network with semi-flexible chains ($\kappa/k_{\text{B}}T=10$), and (iii) entangled network with flexible chains ($\kappa/k_{\text{B}}T=0$), under the same cross-linking fraction $\phi_{\text{c}}=0.1034$. (b) Average pore size $\la d_{\text{pore}} \ra$ as a function of system volume $V$ for regular networks with flexible (dotted lines and open symbols) and semi-flexible strands (solid lines and filled symbols). The light red line shows the results from simulations with weaker interaction strength $\varepsilon_{\text{LJ}}=k_{\text{B}}T/2$.  (c) $\langle d_{\text{pore}} \rangle$ as a function of $V$ for regular (dotted lines and open symbols) and entangled networks (solid lines and filled symbols). The gray line indicates the scaling of $\langle d_{\text{pore}} \rangle \propto V^{1/3}$ observed for regular networks with flexible chains.}
    \label{fig:fig2}
\end{figure*}

To gain an initial qualitative understanding of how network properties influence phase separation, we first ask whether the equilibrium states coarsen indefinitely (macrophase separation) or give rise to domains of finite size (microphase separation). Figure \ref{fig:fig2}(a) shows representative snapshots at the same cross-linking fraction $\phi_{c}$ for the three different networks in a poor solvent: (i) a regular network with flexible chains, (ii) a regular network with semi-flexible chains, and (iii) an entangled networks with flexible chains.
We observed phase separations into monomer-rich and monomer-dilute regions in all three cases, yet the phase separation behaviors are qualitatively different.
In regular networks with flexible strands [Fig. \ref{fig:fig2}(a-i)], the majority of chains collapse into a dense phase, to minimize the interfacial area with the surrounding (implicit) poor solvent.
This reduction in internal energy comes at the expense of a small fraction of chains losing conformational entropy through full stretching, thereby supporting the formation of the large network-dilute phase.
In contrast, increasing the bending rigidity in the regular networks prevents individual chain collapse [Fig. \ref{fig:fig2}(a-ii)], leading instead to the formation of interconnected network with thick nematic bundles.
Note that the cross-linking constraints impede global nematic ordering (see Sec.~S5 in the SI).
In entangled network with flexible chains [Fig. \ref{fig:fig2}(a-iii)], the network-rich phase is composed of collapsed chains, similar to regular networks with flexible chains, but the voids resemble those found in the regular network with semi-flexible chains.

To distinguish between macro- and microphase separation, we performed a finite-size scaling analysis of the average pore diameter, $\langle d_\text{pore} \rangle$. 
The pore diameter $d_\text{pore}$ is defined as the diameter of the largest spherical test probe that can be inserted into the system without overlapping the surrounding monomers (see Sec.~S8 in the SI for technical details).\cite{gelb:lng:1999, bhattacharya:lng:2006, sorichetti:ma:2020}
For macrophase separation, $\langle d_\text{pore} \rangle$ is expected to increase with system size. 
In contrast, for microphase separation, $\langle d_\text{pore} \rangle$ is intrinsic to the network architecture and should therefore remain independent of system size.

Figure \ref{fig:fig2}(b) shows the relationship between the system volume $V$ and the average pore size $\langle d_{\text{pore}} \rangle$ in the regular networks with different chain stiffnesses.
For the flexible case ($\kappa/k_{\text{B}}T = 0$), we find $\langle d_{\text{pore}} \rangle \propto V^{1/3}$, nearly independent of the cross-linking fraction $\phi_{c}$, indicating \textit{macrophase} separation regardless of the chain contour length $L_\text{c}$.
This behavior is preserved even under improved monomer-solvent affinity, though still within the poor-solvent regime (see Sec.~\ref{sec:method} and Sec.~S7 in the SI). Such macrophase separation is consistent with theoretical predictions from the Flory-Huggins model,\cite{qiang:prx:2024} where attractive monomer-monomer interactions drive phase separation once the entropic cost of chain deformation becomes negligible.
In contrast, regular networks with semi-flexible chains ($\kappa/k_{\text{B}}T = 10$) exhibit constant $\langle d_{\text{pore}} \rangle$ for different system sizes -- indicative of \text{microphase} separation -- with higher $\phi_\text{c}$ resulting in smaller $\langle d_{\text{pore}} \rangle$. 
Next, we consider entangled networks with randomly placed cross-links to investigate the role of topological constraints in domain coarsening [Fig. \ref{fig:fig2}(c)].
At low cross-linking fraction ($\phi_\text{c}=0.0508$), the phase separation behavior remains ambiguous, as it is unclear whether the pore size converges with increasing system size.
Above $\phi_\text{c} = 0.0690$, however, $\langle d_{\text{pore}} \rangle$ remains constant beyond a certain system size, with higher $\phi_\text{c}$ producing smaller pores.

These results demonstrate that intrinsic chain stiffness and topological confinement can arrest domain coarsening. However, these two mechanisms differ qualitatively, as indicated by the distinct network morphologies shown in Fig. \ref{fig:fig2}(a-ii) and (a-iii): bending stiffness is a molecular, single-chain property set by the persistence length $L_\text{p}$, whereas topological constraints are mesoscopic and arise from inter-chain entanglements, characterized by the entanglement length $L_\text{e}$. 
We therefore analyze these two distinct pathways to microphase separation separately.

\subsection{Local Bending Rigidity Restricts Domain Coarsening}
\label{subsec:bending_rigidity}
To understand how the bending stiffness of the individual strands modulates the network's properties, we first consider how cross-linking affects the strand conformations. Free flexible polymers in good solvents adopt swollen coil conformations that balance excluded-volume interactions with conformational entropy,\cite{flory:jcp:1949, rubinstein:book:2003} resulting in a root-mean-square radius of gyration that scales as $R_\text{g} \equiv \la R_\text{g}^2 \ra^{1/2} \propto N^{3/5}$ [see Eq.~\eqref{eq:Rg} below].
In contrast, semi-flexible polymers in free solution become increasingly elongated with larger bending stiffness $\kappa$, yet still display pronounced transverse fluctuations even when the persistence length approaches the contour length ($L_{\text{p}}/L_{\text{c}} \sim 1$), reflecting the interplay between bending energy and thermal undulations.\cite{yamakawa:jcp:1973, nikoubashman:jcp:2016}
Similarly, within a network where $L_{\text{c}} \gg a$, the strands also adopt coil-like conformations that elongate as bending rigidity (and consequently the persistence length $L_{\text{p}}$) increases.
When $L_{\text{p}} \gtrsim a$, however, cross-links begin to impose strong geometric constraints, marking a crossover from free-chain behavior to frustrated, network-limited conformations.
Consequently, the effective persistence length $L_\text{p}$ of chains embedded in the network differs from the intrinsic persistence length of an isolated chain $L_{\text{p,0}}$ within this regime (see Fig.~S2 in the SI).

\begin{figure}[htb]
    \centering
    \includegraphics[width=0.9\linewidth]{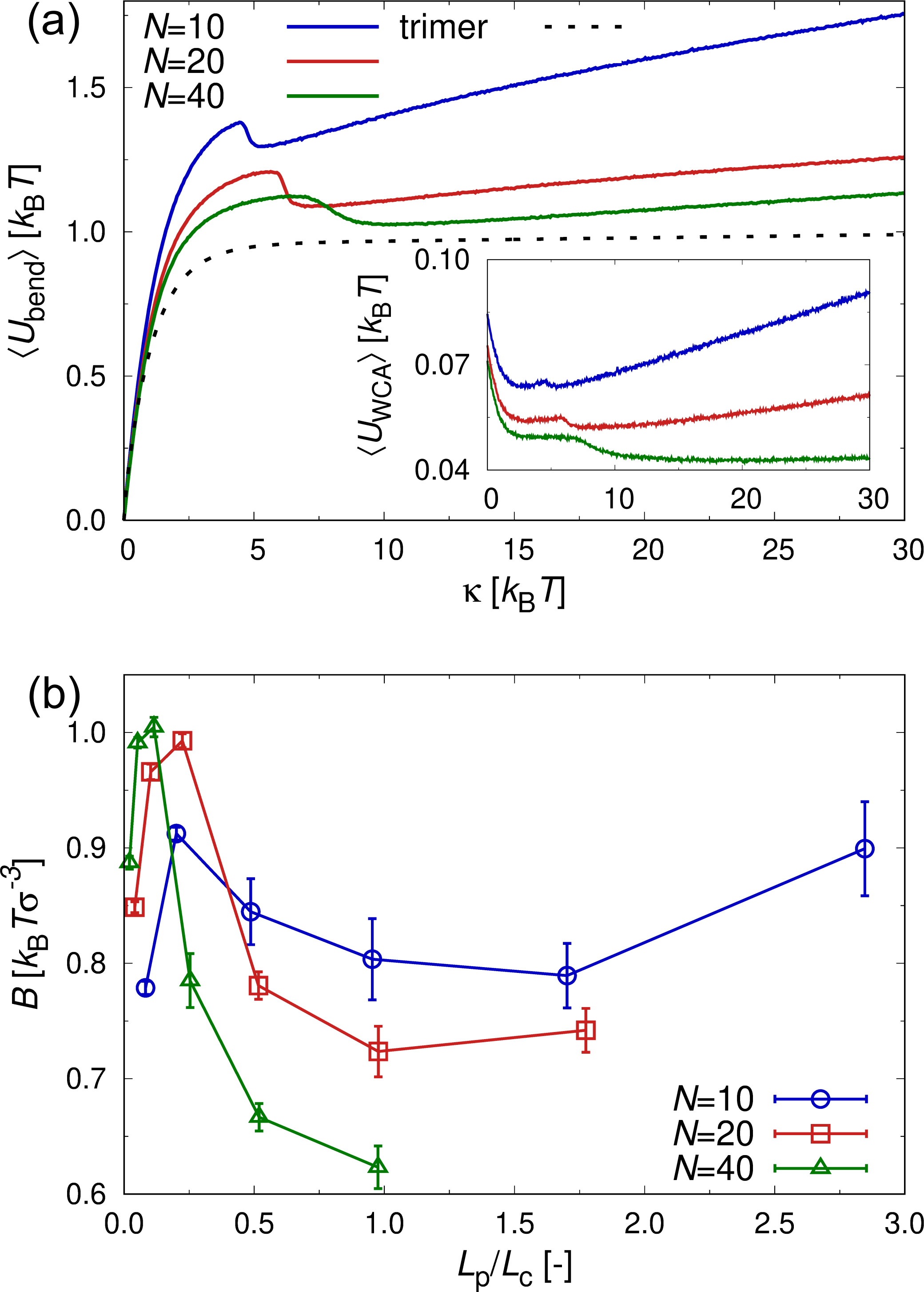}
    \caption{(a) Average bending potential energy per triplet $\langle U_{\text{bend}} \rangle$ (Eq. \eqref{eq:UBEND}) as a function of bending stiffness $\kappa$. Inset: Average monomer-monomer interaction energy per monomer $\langle U_{\text{WCA}} \rangle$ (Eq. \eqref{eq:UWCA}).
    (b) Bulk modulus $B$ of networks as a function of relative persistence length $L_\text{p}/L_\text{c}$. All data shown for regular networks in a good solvent.}
    \label{fig:fig3}
\end{figure}

To assess whether the transition from macro- to microphase separation leaves clear signatures on the molecular level, we analyzed the evolution of the average bending potential energy per triplet $\langle U_{\text{bend}} \rangle$ [Eq. (\ref{eq:UBEND})] with gradually increasing stiffness $\kappa$ under good solvent conditions.
As a reference, we also simulated an isolated trimer with a single bond angle. 
In this case, $\langle U_{\text{bend}} \rangle$ increases linearly for small stiffness parameters $\kappa/k_{\text{B}}T \lesssim 2$ and smoothly approaches the limit of $k_\text{B}T$, consistent with the two quadratic degrees of freedom in the bending potential.
In polymer networks, however, this monotonic behavior breaks down, as shown in Fig.~\ref{fig:fig3}(a). 
At small $\kappa$, $\la U_\text{bend} \ra$ still increases and eventually surpasses $k_\text{B}T$, reflecting frustrated configurations on the segmental level due to the cross-linking constraints.
This initial increase is followed by a sudden drop of the bending energy $\langle U_{\text{bend}} \rangle$ in the region of $4 \lesssim \kappa/k_{\text{B}}T \lesssim 8$, depending on $N$, indicating a relaxation of the (local) chain conformations. 
This drop is accompanied by a sharp increase in the persistence length $L_{\text{p}}$ in the region where $L_{\text{p}}/a \sim 1$ (Fig. S2 in the SI).
Importantly, the $\kappa$-window that exhibits the non-monotonic behavior under good solvent conditions coincides with the cross-over from macro- to microphase separation observed in poor solvents, suggesting that the mismatch between $L_\text{p}$ and $a$ arrests domain coarsening.

As chains straighten with increasing stiffness $\kappa$, monomer-monomer contacts decrease, leading to a drop of the excluded volume interaction energy $\langle U_{\text{WCA}} \rangle$ [Eq. (\ref{eq:UWCA})], as shown in the inset of Fig. \ref{fig:fig3}(a).
This decrease is initially gradual for $\kappa/k_{\text{B}}T \lesssim 2$, followed by a more pronounced drop in the same range where $\langle U_{\text{bend}} \rangle$ falls, reflecting the relaxation of local chain conformations. 
The reduction in short-range monomer-monomer interactions is also consistent with the suppression of correlations in the radial distribution function (Fig. S3 in the SI). 
Together, these energetic signatures indicate that the transition in local chain behavior occurs in the range $4 \lesssim \kappa/k_{\text{B}}T \lesssim 8$ for chain lengths $10 \leq N \leq 40$, with shorter chains requiring lower stiffness to undergo the crossover.

Previous experimental and theoretical studies\cite{fernandez:nmat:2024, qiang:prx:2024, mannattil:prl:2025, oudich:jmps:2026} have related the characteristic phase separation length-scale $\lambda$ to the network elasticity, by considering the network strands as entropic springs. To quantify the overall network elasticity depending on the chain bending stiffness, we computed the bulk modulus $B = \rho \Delta P/\Delta \rho$, by measuring the pressure difference $\Delta P$ after applying $1 \%$ density perturbations ($\Delta \rho$) through expansion and compression under good solvent conditions.
The bulk modulus $B$ can be split up into an ideal term, $B_\text{id} = k_\text{B}T\rho$, which is identical for all investigated systems, and an excess term $B_\text{exc}$ that reflects excluded volume and bonded interactions (see Sec. S10 in the SI). In a good solvent, the contribution from non-bonded interactions are always positive and about $25-50\,\%$ larger than $B_\text{id}$. In contrast, the bonded-contribution can become negative when bonds are extended beyond their equilibrium length (see Fig. S16 in the SI).
We expect that $B$ depends, among others, on the flexibility of its strands, which we quantified through the relative chain rigidity $L_\text{p}/L_\text{c}$: the strands are flexible for $L_{\text{p}}/L_{\text{c}} \ll 1$, and transition from semi-flexible to rod-like when $L_{\text{p}}/L_{\text{c}} \gtrsim 1$.

For small $L_{\text{p}}/L_{\text{c}} \lesssim 1/4$, which corresponds to the regime before the sudden drop in $\langle U_{\text{bend}} \rangle$ [Fig. \ref{fig:fig3}(a)], $B$ increases with relative chain rigidity, reflecting the entropic elasticity of flexible strands.\cite{rubinstein:book:2003} 
As the chain stiffness increases further, the bulk modulus $B$ drops sharply, consistent with the loss in entropic elasticity originating from the reduced conformational degrees of freedom of semi-flexible chains; in the worm-like chain model, the entropic spring constant scales as $L_{\text{p}}^{-1}$, both at small and large deformations, although the finite extensibility introduces strong non-linearities at high strains.\cite{marko:ma:1995} 
With further increasing bending stiffness $\kappa$, the chains become predominantly straight ($L_\text{p}/L_\text{c} > 1$) and the enthalpic contribution from bending rigidity causes $B$ to rise again.
A similar non-monotonic dependence of the shear and Young moduli on bending stiffness has been reported recently for related cross-linked polymer systems,\cite{zheng:eml:2024} corroborating our findings.
These results illustrate that the same bulk modulus $B$ can correspond to vastly different chain conformations, limiting its utility as a single descriptor of the network state.

\begin{figure}[htb]
    \centering
    \includegraphics[width=0.9\linewidth]{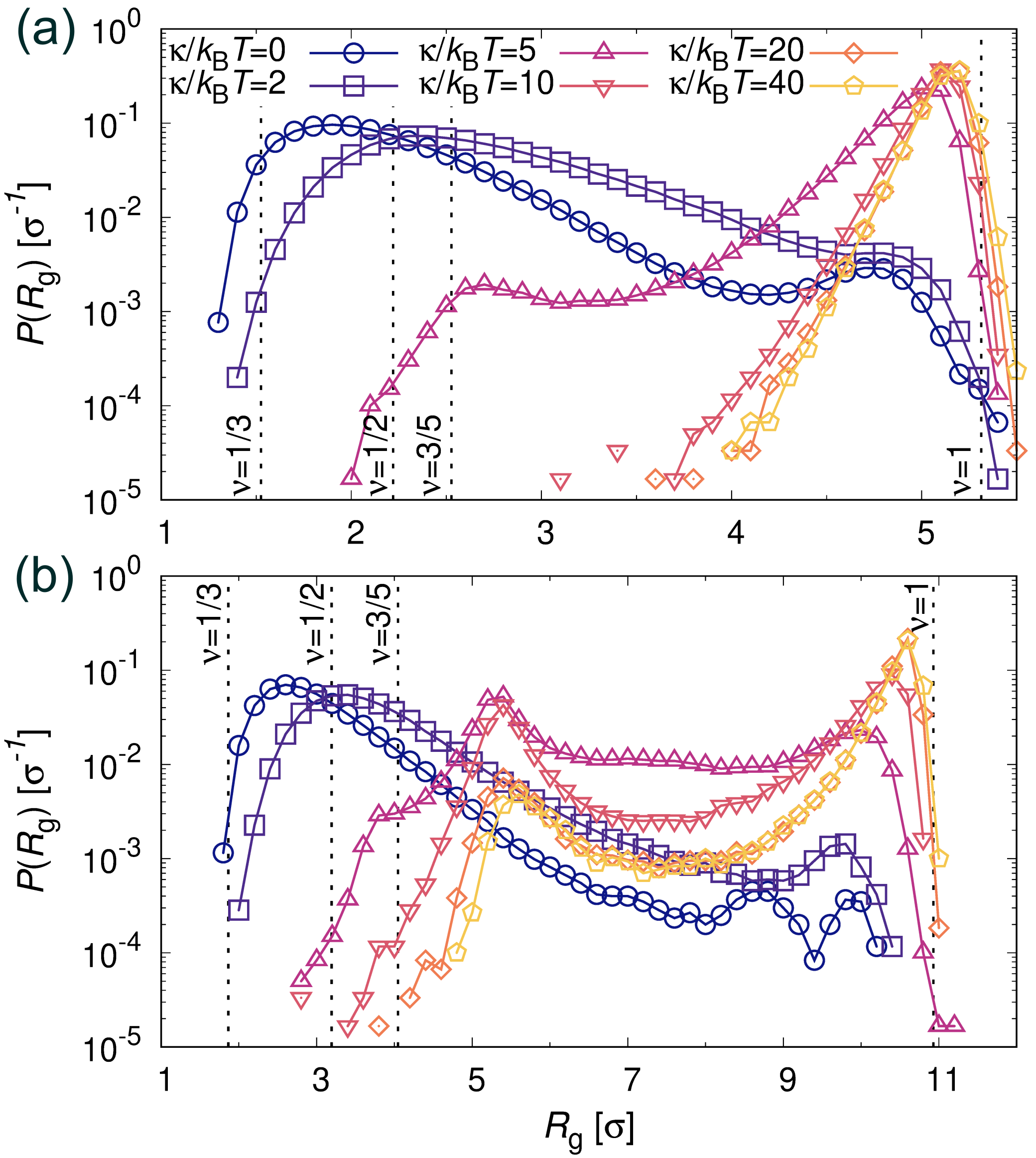}
    \caption{Probability distribution of radius of gyrations $P(R_{\text{g}})$ of network strands for various bending stiffnesses $\kappa$, with chain lengths (a) $N=20$ and (b) $N=40$. Black dotted vertical lines indicate the computed $R_{\text{g}}$ values, corresponding to different chain conformations characterized by their Flory exponents $\nu$: a collapsed globule in a poor solvent ($\nu=1/3$), an ideal random coil in a theta solvent ($\nu=1/2$), a self-avoiding random walk in a good solvent ($\nu=3/5$), and a fully stretched rigid rod ($\nu=1$). The radius of gyration of a rod $R^{\text{rod}}_{\text{g}}$ can be calculated analytically as $R^{\text{rod}}_{\text{g}} = L_{\text{b}} \sqrt{(N^{2} - 1)/12}$. All data shown for regular networks in poor solvent.}
    \label{fig:fig4}
\end{figure}

To connect the phase separation behavior with local structure, we characterized the conformations of network strands between two neighboring cross-linkers through their radius of gyration
\begin{equation}
R_{\text{g}} = \sqrt{\frac{1}{N} \sum_{i=1}^{N} \left(\bm{r}_{i} - \bm{r}_\text{com} \right)^2}
\label{eq:Rg}
\end{equation}
where $\bm{r}_{i}$ and $\bm{r}_{\text{com}}$ denote the position of the $i^{\text{th}}$ monomer, and the center of mass of the strand, respectively.

Figure~\ref{fig:fig4} shows the probability distribution of radius of gyrations $P(R_{\text{g}})$ of network strands in a poor solvent (results for good solvent conditions are shown in Fig. S8 in the SI).
At low stiffness ($\kappa/k_\text{B}T \leq 2$), the distribution $P(R_{\text{g}})$ is rather broad with two local maxima at low and high $R_\text{g}$, indicating the coexistence of collapsed chains in the polymer-rich phase and stretched strands in the dilute phase, respectively [cf. Figs.~\ref{fig:fig2}(a) and \ref{fig:fig4}(a)]. 
This behavior is the \textit{opposite} of that observed for free polymers in solution, which collapse into globules in the dilute phase to minimize exposure to the surrounding poor solvent, and adopt coiled conformations in the polymer-rich phase due to the screening of excluded volume interactions.\cite{kuhn:kz:1934, rubinstein:book:2003}

With increasing stiffness, the population of collapsed chains diminishes drastically, marking the transition from macro- to microphase separation. 
At $\kappa/k_{\text{B}}T = 5$, a small fraction of chains with coil-like conformations remains, but stretched chains dominate. For $\kappa/k_{\text{B}}T \geq 10$, the distribution $P(R_{\text{g}})$ becomes unimodal, indicating that nearly all strands are extended. 
Beyond $L_{\text{p}}/L_{\text{c}} \approx 1/2$, further increases in local bending stiffness $\kappa$ cause little additional chain stretching, due to the topological constraints imposed on the chain conformations by the cross-links.
The mismatch of lattice constant $a$, chain contour length $L_\text{c}$, and persistence length $L_\text{p}$ has additional effects on the chain conformations.
For shorter chains ($N=20$, $L_{\text{c}}/a = 3.69$), most chains are fully stretched at high $\kappa$ (Fig. \ref{fig:fig4}(a)). 
For longer chains ($N=40$, $L_{\text{c}}/a = 5.83$), the contour length far exceeds the lattice spacing, forcing portions of the chains to bend, resulting in the peak of $P(R_{\text{g}})$ at intermediate $R_{\text{g}}$ [Fig. \ref{fig:fig4}(b)].

These transitions in $P(R_{\text{g}})$ reflect the balance between monomer-monomer attraction and chain rigidity for the phase behavior in regular networks.
For flexible strands, attractive interactions dominate, causing most chains to form a large polymer-rich phase, at the expense of few fully stretched chains in the dilute phase, leading to macrophase separation.
Once the chains are sufficiently stiff to resist collapse, microphase separation emerges in the form of nematic bundles.

\begin{figure}[htb]
    \centering
    \includegraphics[width=0.90\linewidth]{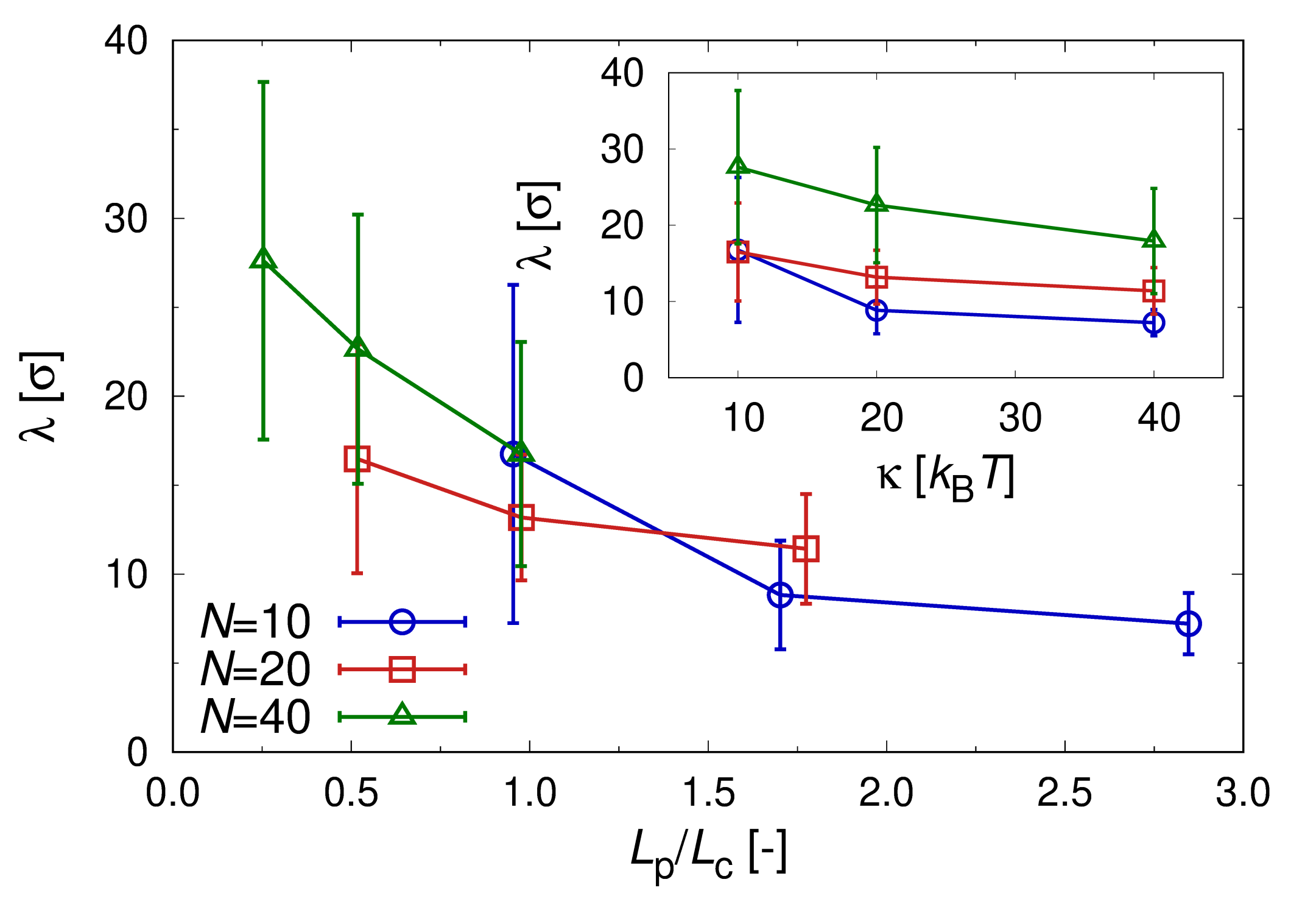}
    \caption{Relationships between the characteristic length scale $\lambda$ of microphase separation an relative chain rigidity $L_{\text{p}}/L_{\text{c}}$ in regular networks for various chain lengths $N$. Inset: $\lambda$ as a function of stiffness parameter $\kappa$. Error bars indicate the standard deviation of $\lambda$ within one system.}
    \label{fig:fig5}
\end{figure}

To identify which network-specific length scales control the microphase separation length $\lambda$, defined as the average pore size, $\lambda =  \langle d_\text{pore} \rangle$, we focus on networks with $\kappa/k_{\text{B}}T \geq 10$, where bending rigidity suppresses domain coarsening.
Natural candidates are the persistence length $L_\text{p}$, which determines the stiffness of individual strands, and the contour length $L_\text{c}$, which (indirectly) reflects the distance between neighboring cross-links that constrain possible chain conformations.
Their ratio captures the interplay between local chain mechanics and network topology, and should directly regulate the emergent length scale $\lambda$. 
Indeed, flexible strands form larger pores, with $\lambda$ decreasing almost linearly with increasing $L_\text{p}/L_\text{c}$ for $L_\text{p}/L_\text{c} \lesssim 1$, independent of $N$ (Fig.~\ref{fig:fig5}). 
As the strands become even stiffer, the pore size becomes almost independent of $L_\text{p}/L_\text{c}$, reflecting the conformational constraints imposed by the cross-links.
The inset of Fig. \ref{fig:fig5} shows that shorter strands lead to smaller pores $\lambda$ at the same stiffness parameter $\kappa$, confirming that both bending rigidity and contour length contribute through their ratio.

\subsection{Topological Constraints Restrict Domain Coarsening}
\label{subsec:topological_constraints}
As already indicated in Sec.~\ref{subsec:phase_separation}, microphase separation can also be achieved by imposing topological constraints.
To examine these effects in more detail, we systematically study in this section the relations between the microscopic entanglement length $L_\text{e}$, and the resulting elastic properties and phase separation behavior of the polymer network.
In entangled networks, the initial chain contour length is fixed at $L_{\text{c}} = 37.83\,\sigma$, and we tuned the density of topological constraints by varying the cross-linking fraction and also by applying a temporary bending potential to straighten chains prior to cross-linking (see Sec.~\ref{sec:method} for details). 
Higher pre-cross-linking stiffness yields more entangled networks, but this procedure is restricted to the isotropic regime prior nematic ordering (the auxiliary bending potential was removed after cross-linking).

We quantify the degree of topological constraints through the average entanglement length $L_\text{e}$, defined as
\begin{equation}
    L_{\text{e}} = \frac {L_\text{c}}{\langle Z \rangle}
    \label{eq:Ne}
\end{equation}
where $L_\text{c}$ is the contour length before cross-linking, and $\langle Z \rangle$ is the average number of kinks identified along the primitive path of polymer strands (the shortest path connecting the two chain ends while preserving the chain topology).
Each kink corresponds to an entanglement point, and they were identified via primitive path analysis using the Z1+ algorithm.\cite{kroeger:cpc:2023}
We calculated the entanglement length $L_{\text{e}}$ under good solvent conditions to characterize the network state prior to phase separation.
Note that Eq.~\eqref{eq:Ne} is only reliable for sufficiently entangled systems with $L_{\text{e}} \lesssim L_{\text{c}}$.\cite{hoy:pre:2009}

\begin{figure}[htb]
    \centering
    \includegraphics[width=0.90\linewidth]{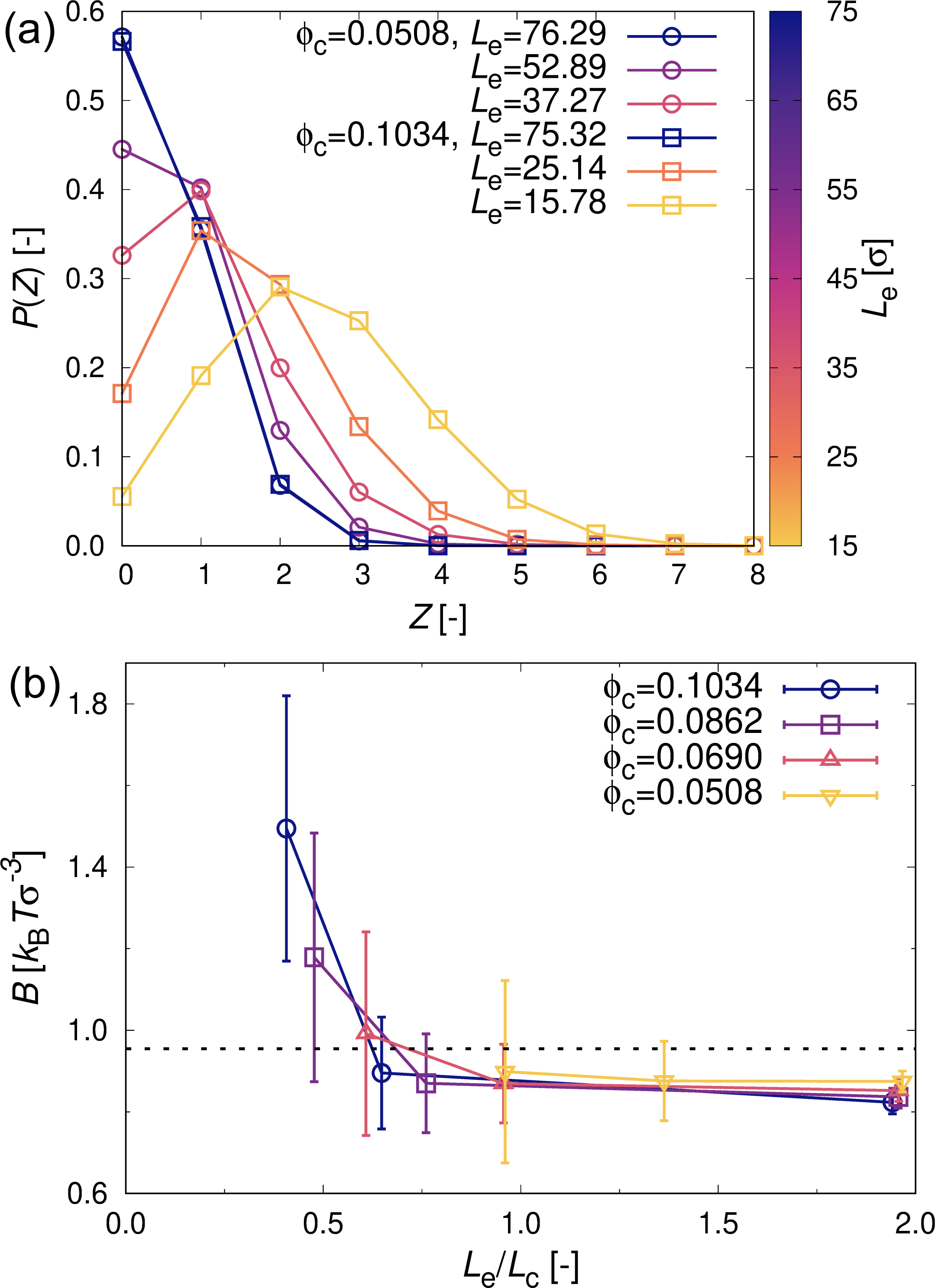}
    \caption{(a) Probability distribution of number of entanglement points along the chain contour, $P(Z)$, for different cross-linker fractions $\phi_\text{c}$ and average entanglement lengths $L_{\text{e}}$. (b) Bulk modulus $B$ of entangled networks in a good solvent as a function of $L_{\text{e}}/L_{\text{c}}$, at fixed contour length $L_{\text{c}} \approx  37.83\,\sigma$ for all entangled networks. The dashed horizontal line indicates $B$ before cross-linking.}
    \label{fig:fig6}
\end{figure}

First, we examined how entanglements are distributed across strands in the networks.
Figure \ref{fig:fig6}(a) shows the probability distribution of the number of entanglement points per chain, $P(Z)$, in good solvents, resolved for different $\phi_\text{c}$ and $L_{\text{e}}$.
In weakly entangled networks ($L_{\text{e}} > L_{\text{c}}$), more than half of the strands are free of entanglements ($Z=0$).
Even in systems, where the average entanglement length is comparable to the contour length ($L_{\text{e}} \approx L_{\text{c}}$), about one third of the strands remain unentangled.
Comparing networks with similar $L_{\text{e}} \approx 76\,\sigma$ reveals nearly identical $P(Z)$, irrespective of $\phi_\text{c}$.
As $L_{\text{e}}$ decreases, the probability of chains carrying multiple entanglements ($Z \ge 1$) increases, demonstrating that $L_{\text{e}}$ provides a faithful descriptor of topological constraints, whereas $\phi_\text{c}$ alone does not.

Next, we studied how topological constraints influence the elastic modulus $B$ of the polymer networks under good solvent conditions [Fig. \ref{fig:fig6}(b)].
The bulk modulus $B$ increases with decreasing $L_{\text{e}}$, indicating that networks with more entanglements (shorter $L_{\text{e}}$) are stiffer.
Networks with similar $L_{\text{e}}$ have comparable $B$, regardless of their cross-linking fraction $\phi_\text{c}$, confirming that $L_{\text{e}}$ is a robust indicator of the network’s elasticity.
In the weakly entangled regime ($L_{\text{e}}/ L_{\text{c}} > 1$), $B$ shows little variation with $L_\text{e}$, consistent with the small changes in $P(Z)$ in that regime. Notably, $B$ in the weakly entangled networks is smaller than the value before cross-linking [dashed horizontal line in Fig.~\ref{fig:fig6}(b)]. This reduction arises because cross‑linking increases the number of bonded monomers, thereby reducing the positive contribution to $B_\text{exc}$ from the purely repulsive, excluded‑volume interactions among non-bonded monomers (see Fig. S17 in the SI).

\begin{figure}[htb]
    \centering
    \includegraphics[width=0.90\linewidth]{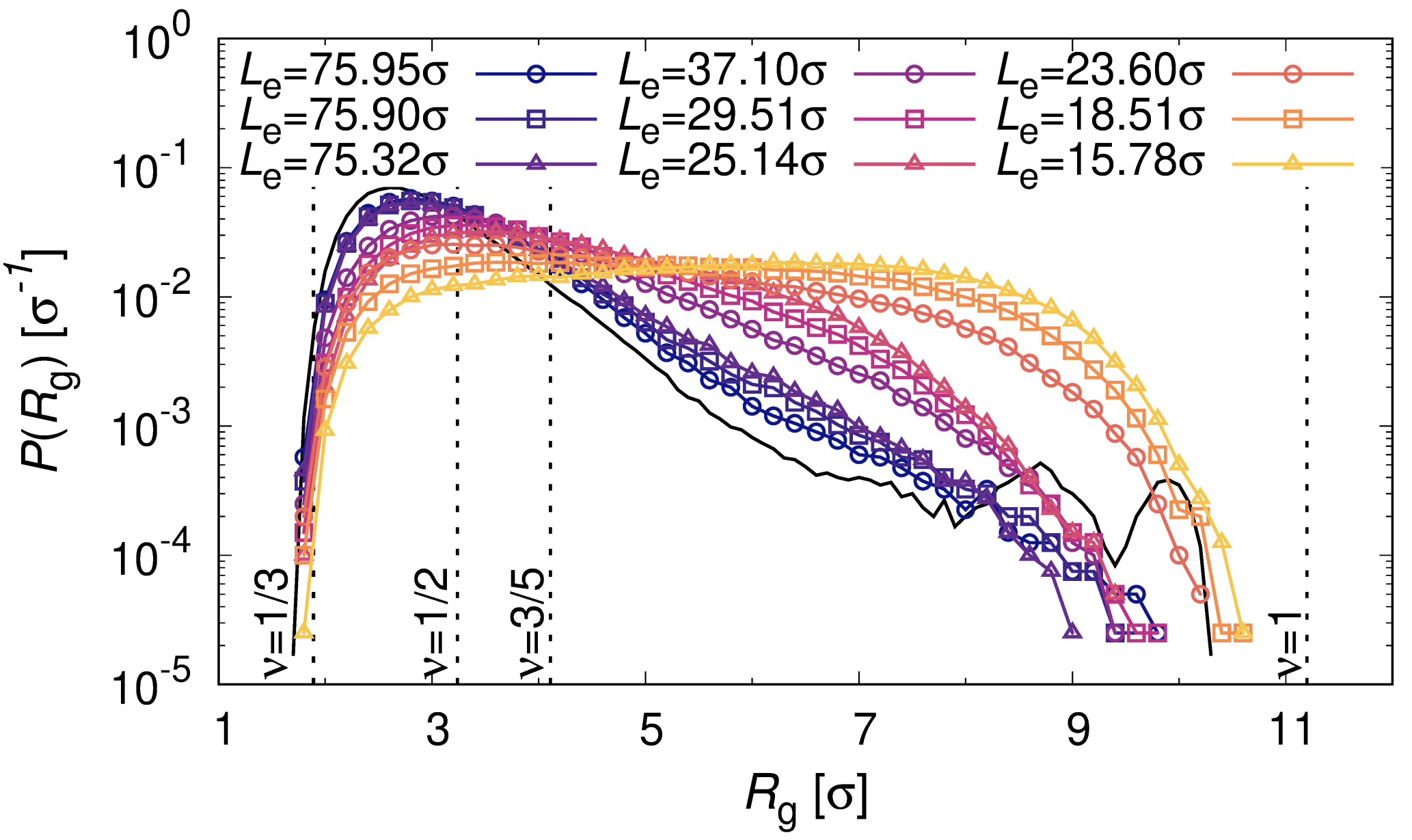}
    \caption{Distribution of the radius of gyration $P(R_{\text{g}})$ of chains in microphase-separated entangled networks with different entanglement lengths $L_{\text{e}}$. The circle, square, and triangle symbols correspond to cross-linking fractions $\phi_\text{c}=0.0690,\ 0.0862, \ 0.1034$, respectively. The black curve shows $P(R_{\text{g}})$ from a regular, macrophase-separated network with flexible chains of length $N=40$. Black dotted vertical lines indicate the computed $R_{\text{g}}$ values corresponding to different chain conformations characterized by their Flory exponents $\nu$: a collapsed globule in a poor solvent ($\nu=1/3$), an ideal random coil in a theta solvent ($\nu=1/2$), a self-avoiding random walk in a good solvent ($\nu=3/5$), and a fully stretched rigid rod ($\nu=1$).}
    \label{fig:fig7}
\end{figure}

Unlike bending stiffness, which primarily acts at the single-chain level, entanglements emerge from multi-chain interactions, making their effect on chain conformations less direct. 
To elucidate these effects, we computed $P(R_\text{g})$ for microphase-separated entangled networks ($\phi_{c} \ge 0.0690$) realized at different $L_{\text{e}}$ (Fig.~\ref{fig:fig7}). 
Interestingly, the chain conformations in weakly entangled networks are similar to those of the macrophase-separated regular network without topological entanglements (black line in Fig.~\ref{fig:fig7}). 
With decreasing entanglement length $L_\text{e}$, $P(R_\text{g})$ broadens markedly, which is in stark contrast to the much more narrow $P(R_\text{g})$ observed in semi-flexible microphase-separated networks (cf. Fig.~\ref{fig:fig4}).
These findings indicate that the phase separation behavior of entangled networks cannot be inferred from the conformations of individual polymer strands; instead, it is governed by the collective length scale $L_\text{e}$ associated with chain interactions.
In other words, physical constraints imposed by the network topology, rather than strand-specific features, prevent most chains from collapsing into a single dense domain.

\begin{figure}[htb]
    \centering
    \includegraphics[width=0.90\linewidth]{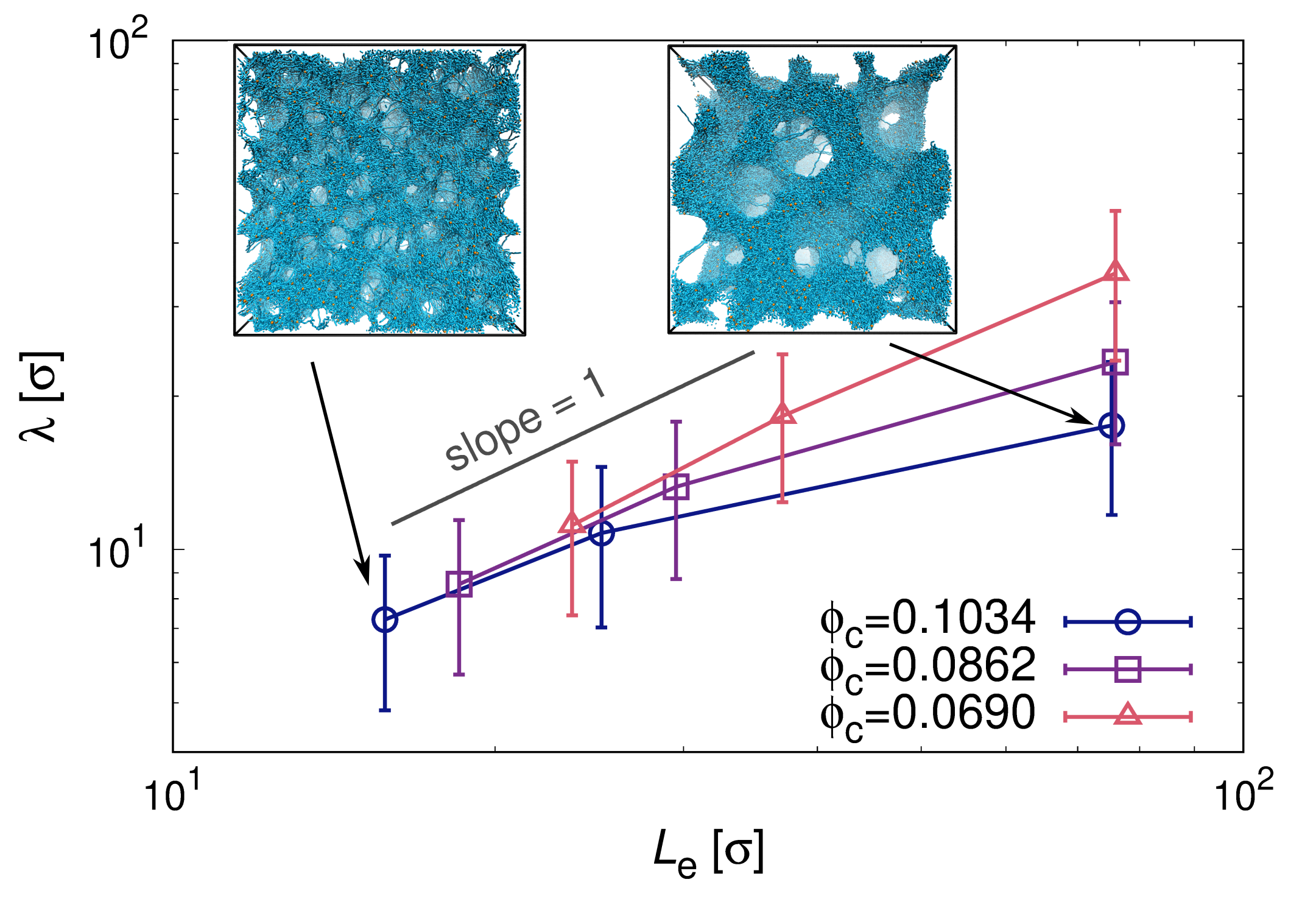}
    \caption{Characteristic length scale of microphase separation $\lambda$ as a function entanglement length $L_{\text{e}}$ for entangled networks with different cross-linking fraction $\phi_\text{c}$. The snapshots show typical network configurations, as indicated. Error bars indicate the standard deviation of $\lambda$ within one system.}
    \label{fig:fig8}
\end{figure}

As proposed in recent field-theory work,\cite{qiang:prx:2024} a non-local length scale is key for achieving microphase separation in elastic polymer networks.
To test whether this length scale originates from topological constraints, we examined the relationship between the entanglement length $L_{\text{e}}$ and the characteristic length scale of microphase separations $\lambda$ (Fig. \ref{fig:fig8}).
In the highly entangled regime ($L_{\text{e}} \lesssim L_{\text{c}}$), we find $\lambda \propto L_{\text{e}}$, with data from networks of different cross-linking fractions $\phi_\text{c}$ collapsing onto a same scaling line.  
This growth of $\lambda$ is accompanied by only a slight reduction in $B$; the correlation is weak, and in weakly entangled networks $\lambda$ is nearly independent of $B$ (see Fig. S18 in the SI).
This behavior suggests that the phase separation length scale $\lambda$ is primarily governed by the entanglement between multiple strands rather than by the entropic elasticity of individual strands.
Thus, the non-local length scale posited by field theory can be attributed to the characteristic length scale from topological constraints, which emerge from multi-chain interactions.
This perspective also explains why the size of microphase-separated domains in flexible-chain networks can exceed the typical end-to-end distance of individual strands.\cite{fernandez:nmat:2024} 

\section{Conclusions}
Polymer networks are central to the structure and mechanics of both synthetic and living materials. A key feature of these systems is their ability to arrest phase separation of embedded immiscible components, with domain sizes that often exceed the network mesh size. Recent theories have suggested that such limited domain growth requires a network‑specific length scale, but its microscopic physical origin has remained elusive. Using molecular simulations, we built three model networks to isolate candidate scales: (i) a regular network with flexible strands, (ii) a regular network with semi-flexible strands, and (iii) an entangled network with flexible strands. By performing a careful finite‑size scaling analysis, we found that regular flexible networks undergo macrophase separation. In contrast, microphase separation emerged when the persistence length becomes comparable to the cross‑link spacing, because single-chain bending stiffness resists the collapse into a dense phase. Irregular networks of flexible polymers can also arrest domain coarsening when strands are sufficiently entangled, owing to the additional topological constraints imposed at the mesoscale level. In both cases, the resulting microphase spacing is much larger than the typical strand dimensions in a good solvent, consistent with experimental observations. Finally, we observed that the bulk elasticity does not reliably predict phase behavior, as we found systems with similar bulk modulus that differ qualitatively in phase separation behavior. An entropic‑spring description is therefore insufficient; chain stiffness and network topology (entanglements) must be included to capture the relevant length scales that arrest coarsening.

Looking ahead, two biological relevant extensions are particularly compelling. First, strand activity could provide an independent arrest mechanism due to the competition between contractile stresses, interfacial tension and network elasticity. Second, wetting and adhesion to the network could redirect phase separation from immersed droplets to coatings on the network filaments, with contact-line pinning and partial wetting governed by condensate-network affinity.

\section{Models and Methods}
\label{sec:method}
\subsection{Molecular Dynamics}
We use a coarse-grained bead-spring model, where polymer chains are represented as sequences of spherical monomeric units with diameter $\sigma$ and unit mass $m$.
The solvent is modeled implicitly, and the solvent quality is incorporated into the effective monomer-monomer interaction.
Excluded volume interactions in a good solvent are modeled through the purely repulsive Weeks-Chandler-Andersen (WCA) pair potential\cite{weeks:jcp:1971}
\begin{equation}
    U_\text{WCA}(r) = 
    \begin{cases}
    4\varepsilon\left[\left( \sigma /r\right)^{12} - \left( \sigma /r\right)^6 + 1/4 \right],& r \leq 2^{1/6}\,\sigma \\
    0, & r > 2^{1/6}\,\sigma
    \end{cases}
    \label{eq:UWCA}
\end{equation}
where $r$ is the distance between two monomers, and $\varepsilon$ controls the interaction strength.
The fundamental units in the model are defined in terms of $\sigma$, $m$, and $\varepsilon$, corresponding to length, mass, and energy, respectively.
Under poor solvent conditions, the solvent-mediated attraction between monomers is described by the Lennard-Jones (LJ) potential, 
\begin{equation}
    U_\text{LJ}(r) = 
    \begin{cases}
    4\varepsilon_{\text{LJ}}\left[\left( \sigma /r\right)^{12} - \left( \sigma /r\right)^6 \right] ,& r \leq 2.5 \sigma \\
    0, & r > 2.5 \sigma
    \end{cases},
    \label{eq:ULJ}
\end{equation}
with $\varepsilon_{\text{LJ}}$ being the interaction strength of the LJ potential.

Monomers are bonded via the finitely extensible nonlinear elastic (FENE) potential,\cite{bishop:jcp:1979}
\begin{equation}
    U_\text{FENE}(r) = 
    \begin{cases}
  -\frac{1}{2}k r_{0}^2  \ln \left[ 1 - \left(r/r_{0} \right)^2 \right], & r < \,r_{0}\\
    \infty, & r \geq r_{0}
    \end{cases} .
    \label{eq:UFENE}
\end{equation}
with maximum bond extension $r_0 = 1.5\,\sigma$ and spring constant $k=30\, k_{\text{B}}T/\sigma^{2}$, resulting in an equilibrium bond length of $L_{\text{b}} \approx 0.97\,\sigma$.
These parameters are chosen to avoid unphysical bond crossing.\cite{grest:pra:1986}

Bending stiffness of semi-flexible chains is implemented via the harmonic bending potential
\begin{equation}
    U_{\text{bend}}(\Theta_{ijk}) = \frac {\kappa}{2}  \Theta_{ijk}^{2},
    \label{eq:UBEND}
\end{equation}
where $\kappa$ controls the (local) bending stiffness, and $\Theta_{ijk}$ is the angle between two subsequent vectors, $\bm{r}_{ij}$ and  $\bm{r}_{jk}$ connecting the monomers $i$, $j$, and $k$ within a chain.
Thus, the potential $U_{\text{bend}}(\Theta_{ijk})$ vanishes when monomers $i$, $j$, and $k$ lie on a straight line. 

All molecular dynamics (MD) simulations are performed in the $\mathcal{N}VT$ ensemble using the HOOMD-blue software package (v. 4.7.0),\cite{anderson:cms:2020} where $\mathcal{N}$ is the total number of monomers in the system, and $V$ is the volume of the cubic simulation box.
Periodic boundary conditions are applied to all three Cartesian directions.
Unless otherwise specified, all simulations are performed at a constant temperature $k_\text{B}T=\varepsilon$ by using a Langevin thermostat.
The friction coefficient is set to $\gamma = m/\tau$ with $\tau = \sqrt{m \sigma^2/\varepsilon}$ being the intrinsic MD unit of time.
The equations of motion are solved using a velocity Verlet integration scheme with time step $\Delta t = 0.005\,\tau$. 

\subsection{Simulation Details}
All polymer networks are first prepared under good solvent conditions.
To equilibrate the regular networks with flexible chains, we started from configurations without monomer overlap and ran MD simulations for $5 \times 10^{3} \tau$, which was sufficient to reach a steady potential energy.
Networks composed of semi-flexible chains were initialized from equilibrated the fully flexible networks, and we slowly increased the bending stiffness $\kappa$ at a rate of $1/3 \times 10^{-5} \kappa / \tau$ until the desired value of $\kappa$ was reached at an increased temperature $k_{\text{B}}T = 2\varepsilon$.
This step was followed by cooling down to $k_{\text{B}}T = \varepsilon$ over $5 \times 10^{5} \tau$, resulting in the final state of semi-flexible chain networks in a good solvent.
Despite these efforts, equilibrating the semi-flexible chain networks was considerably more challenging due to the mismatch in length scales, i.e., $a \ll L_{\text{c}}$ and $a \lesssim L_{\text{p}}$ (Fig. \ref{fig:fig1}(a-3)); when $a$ is much smaller than both $L_{\text{c}}$ and $L_{\text{p}}$, the network-imposed constraints force filaments into energetically unfavorable configurations. 
Moreover, the relaxation of the chain conformations is further constrained by the surrounding and connected rigid chains.
Such slow coarsening behavior in semi-flexible chain networks has been reported by others as well, and achieving full equilibration within typical simulation timescales remains difficult.\cite{kroeger:sm:2025}
For the entangled networks with flexible chains ($\kappa/k_{\text{B}}T = 0$), we ran MD simulations in good solvent for $5 \times 10^{3} \tau$ for both before and after the cross-linking procedure, to ensure that the linear polymer chains are relaxed within the networks.
To control the degree of topological constraints within the network, the harmonic bending potential (Eq. \ref{eq:UBEND}) was applied to the linear chains prior to the cross-linking procedure.
The bending rigidity parameter $\kappa$ was selected arbitrarily within the range where the polymer solutions exhibited no nematic ordering behavior (see SI for additional details).
This temporal bending rigidity was removed after the cross-linking.

To trigger the phase separation, we changed the solvent quality from good to poor (corresponding to cooling down the systems with upper critical solution temperatures) by changing the monomer-monomer interactions from the WCA potential (Eq. \ref{eq:UWCA}) to the LJ potential (Eq. \ref{eq:ULJ}).
Here, we switched the potential from the WCA to the LJ potential instantly but with a weak LJ interaction strength of $\varepsilon_{\text{LJ}}=\varepsilon/10$ and gradually increased it to $\varepsilon_{\text{LJ}} = \varepsilon$ over $4.5 \times 10^{5} \tau$ to avoid sudden changes in the solvent quality.

To determine the geometric pore sizes $d_\text{pore}$ in the polymer networks, we first created a coarse representation on a cubic grid with cell size $(2.5\,\sigma)^3$. 
Here, we assigned cells with local monomer density $\rho < 0.6\,\sigma^{-3}$ as empty and as full otherwise. 
hen, we placed 50,000 spherical test probes at random positions $\mathbf{r}$, and optimized the position and diameter of each spherical probe such that it still encompasses the original insertion point $\mathbf{r}$ without overlapping the surrounding filled cells (see SI for additional details).\cite{gelb:lng:1999, bhattacharya:lng:2006, sorichetti:ma:2020}

All the analyses of polymer networks except the finite-size scaling analysis were conducted with the system size of $V \approx (120\sigma)^{3}$ to ensure that the side length of the cubic box $L$ is always at least three times longer than the chain contour length $L_{\text{c}}$ (see SI for additional details).

\section*{Data availability}
The data that support the findings of this study are available from the corresponding authors upon reasonable request.

\section*{Conflicts of interest}
There are no conflicts to declare.

\section*{Acknowledgements}
We thank Chengjie Luo and Oliver Paulin for helpful discussions. This work was funded by the Deutsche Forschungsgemeinschaft (DFG, German Research Foundation) through Project No. 470113688, and received support by the Klaus Tschira Foundation through Project No. 00.050.2024/ID 25347.
D.Z. and Y.Q. gratefully acknowledge funding from the Max Planck Society and the European Union (ERC, EmulSim, 101044662). 
The authors acknowledge the Center for High-Performance Computing (ZIH) Dresden for providing computational resources.

\providecommand*{\mcitethebibliography}{\thebibliography}
\csname @ifundefined\endcsname{endmcitethebibliography}
{\let\endmcitethebibliography\endthebibliography}{}

\end{document}


\title{\Large Supplementary Information: Molecular simulations of phase separation in elastic polymer networks}

\author{Takahiro Yokoyama}
 \affiliation{Leibniz-Institut f{\"u}r Polymerforschung Dresden e.V., Hohe Stra\ss e 6, 01069 Dresden, Germany}
 \affiliation{Institut f{\"u}r Theoretische Physik, Technische Universit{\"a}t Dresden, 01069 Dresden, Germany}

\author{Yicheng Qiang}
 \affiliation{Max Planck Institute for Dynamics and Self-Organization, Am Fa\ss berg 17, 37077 G{\"o}ttingen, Germany}

\author{David Zwicker}
\email{david.zwicker@ds.mpg.de}
 \affiliation{Max Planck Institute for Dynamics and Self-Organization, Am Fa\ss berg 17, 37077 G{\"o}ttingen, Germany}

\author{Arash Nikoubashman}
 \email{anikouba@ipfdd.de}
 \affiliation{Leibniz-Institut f{\"u}r Polymerforschung Dresden e.V., Hohe Stra\ss e 6, 01069 Dresden, Germany}
 \affiliation{Institut f{\"u}r Theoretische Physik, Technische Universit{\"a}t Dresden, 01069 Dresden, Germany}
 \affiliation{Cluster of Excellence Physics of Life, Technische Universit{\"a}t Dresden, 01062 Dresden, Germany}
 
\maketitle

\tableofcontents
\clearpage

\section{Preparation of Polymer Networks}
To prepare the regular polymer networks, we first place the monomers on a simple cubic lattice, which results in considerable overlap between the monomers. To eliminate this overlap, we use a purely repulsive Gaussian pair potential for the non-bonded interactions
\begin{equation}
    U_\text{g}(r) = \varepsilon_{\text{g}} \exp \left[ - \left( r / \sigma \right)^2 \right],
    \label{eq:UGAUSS}
\end{equation}
with interaction strength $\varepsilon_\text{g} = 200\,k_\text{B}T$. Furthermore, the monomers are bonded through harmonic springs with equilibrium bond length $\sigma$ and spring constant $500\,k_\text{B}T/\sigma^2$. After eliminating the monomer overlaps, we replaced these two soft potentials with their harder counterparts $U_\text{WCA}$ and $U_\text{FENE}$, respectively (see main text).

The randomly entangled networks were created by placing hexa-functional cross-linkers and linear chains at random positions in the simulation box. After removing the initial monomer overlap using the soft potentials, we equilibrated the system  under good solvent conditions for $5 \times 10^3\,\tau$. To construct the networks, additional bonds were introduced between free cross-linkers and monomers from the linear chains. To ensure network percolation and to avoid the presence of unconnected free polymer chains, one cross-linking point was first assigned to each linear chain. Next, the remaining free cross-linkers were connected to their nearest monomers. These cross-linking bonds were initially modeled as harmonic springs with equilibrium bond length $\sigma$. The spring constant was gradually increased from $20\,k_\text{B}T/\sigma^{2}$ to $500\,k_\text{B}T/\sigma^{2}$ over $5 \times 10^{3} \,\tau$ to avoid sudden local deformations. After this stabilization period, the harmonic springs were replaced by FENE bonds, and the systems were simulated for $5 \times 10^3\,\tau$ to relax the chain conformations.

All simulations were conducted in the $\mathcal{N}VT$ ensemble, with total number of particles $\mathcal{N}$ and volume of the cubic simulation box $V = L^{3}$ (see Tables \ref{tbl:tbl1}-\ref{tbl:tbl4} below). The monomer density was fixed to $\rho \equiv \mathcal{N}/V = 0.4\,\sigma^{-3}$, and the temperature was set to $k_\text{B}T=\varepsilon$ using a Langevin thermostat, unless explicitly stated otherwise.

\begin{table}[tb]
\small
\caption{Simulation details for regular networks with $N=10$.}
  \begin{tabular*}{0.40\textwidth}{@{\extracolsep{\fill}}cc}
    \hline
    $L$ [$\sigma$] & $\mathcal{N}$ \\
    \hline
    41.213 & 28,000 \\
    61.819 & 94,500 \\
    78.304 & 192,052 \\
    98.911 &  387,072 \\
    119.517 & 682,892 \\
    148.366& 1,306,368 \\
    201.943 & 3,294,172 \\
    \hline 
    \end{tabular*}
    
    \label{tbl:tbl1}
\end{table}

\begin{table}[tb]
\small
\caption{Simulation details for regular networks with $N=20$.}
  \begin{tabular*}{0.40\textwidth}{@{\extracolsep{\fill}}cc}
    \hline
    $L$ [$\sigma$] & $\mathcal{N}$ \\
    \hline
    42.029 & 29,696 \\
    57.789 & 77,198 \\
    78.804 & 195,750 \\
    99.818 &  397,822 \\
    120.833 & 705,686 \\
    152.354 & 1,414,562 \\
    199.636 & 3,182,576 \\
    \hline
    \end{tabular*}
    
    \label{tbl:tbl1}
\end{table}

\begin{table}[tb]
\small
\caption{Simulation details for unentangled networks with $N=40$.}
  \begin{tabular*}{0.40\textwidth}{@{\extracolsep{\fill}}cc}
    \hline
    $L$ [$\sigma$] & $\mathcal{N}$ \\
    \hline
    39.942 & 25,488 \\
    59.912 & 86,022 \\
    79.883 & 203,904 \\
    99.854 &  398,250 \\
    119.825 & 688,176 \\
    153.109 & 1,435,706 \\
    199.708 & 3,186,000 \\
    \hline
    \end{tabular*}
  \label{tbl:tbl3}
\end{table}

\begin{table}[tb]
\small
\caption{Simulation details for entangled networks composed of linear chains (40-mer) and hexa-functional cross-linkers.}
  \begin{tabular*}{0.60\textwidth}{@{\extracolsep{\fill}}ccc}
    \hline
    $L$ [$\sigma$] & $\mathcal{N}$ & Total number of 40-mers\\
    \hline
    41.134 & 27,840 & 612\\
    60.405 & 88,160 & 1,938\\
    80.218 & 206,480 & 4,539\\
    120.009 & 691,360 &15,198\\
    200.033 & 3,201,600 & 70,380\\
    \hline
    \end{tabular*}
  \label{tbl:tbl4}
\end{table}

\clearpage
\section{Slow Dynamics of Semi-flexible Chain Networks}
\label{sec:slow_dynamics}
\begin{figure}[htb]
    \centering
    \includegraphics[width=1.00\linewidth]{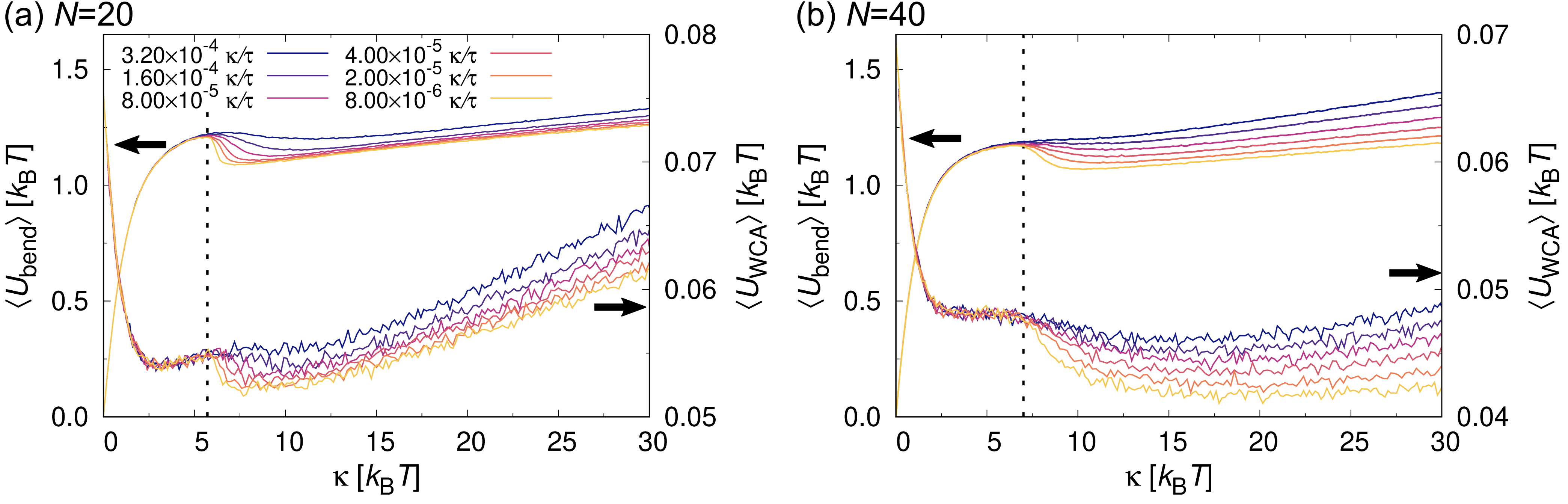}
    \caption{Average bending potential energy per triplet $\la U_\text{bend} \ra$ (left $y$-axis) and average excluded volume interaction energy per monomer $\la U_\text{WCA} \ra$ (right $y$-axis) vs. bending stiffness $\kappa$. Data shown for regular networks with (a) $N = 20$ and (b) $N = 40$ at different increase rates of $\kappa$, as indicated.}
    \label{fig:increase-speed}
\end{figure}

To prepare polymer networks with semi-flexible chains, we started from flexible chain networks and gradually increased the bending stiffness $\kappa$ to its desired value. Figure~\ref{fig:increase-speed} shows the evolution of the bending energy (per triplet), $\la U_\text{bend} \ra$, and the energy due to excluded volume interactions, $\la U_\text{WCA} \ra$, for different rates of increase of $\kappa$. Up to a certain $\kappa$ value, the potential energies change identically for the different increase rates; this threshold roughly corresponds to the point where the persistence length becomes identical to the lattice spacing, $L_\text{p} \approx a$. Beyond this point, the average potential energies become lower when $\kappa$ is increased more slowly.
In practice, it was challenging to reach a sufficiently slow increase rate such that the potential energy evolution became completely independent of the rate.

\clearpage
\section{Chain Expansion and Structural Evolution with Increasing Bending Rigidity}
\begin{figure}[htb]
    \centering
    \includegraphics[width=1.00\linewidth]{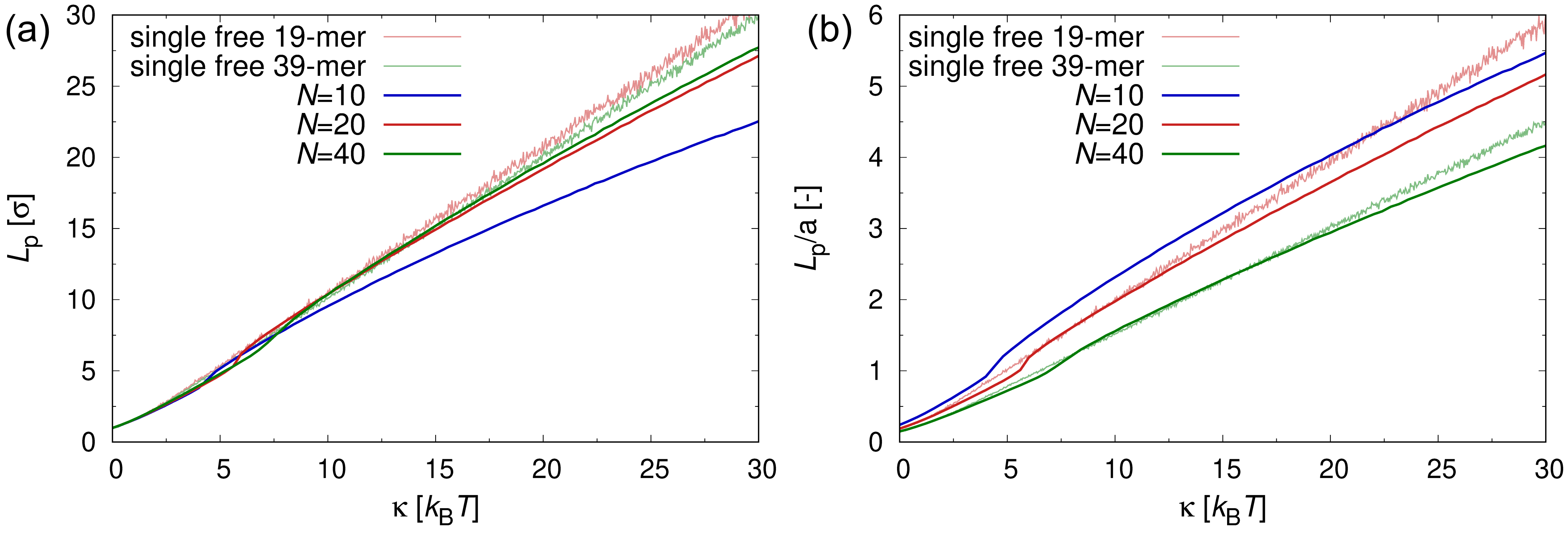}
    \caption{(a) Evolution of persistence length $L_\text{p}$ as a function of $\kappa$. The dashed lines show the behavior of free semi-flexible chains in solution. (b) Evolution of $L_\text{p}$ normalized by the lattice constant $a$ as a function of $\kappa$.}
    \label{fig:evolution_lp}
\end{figure}

During the preparation of unentangled networks with semi-flexible chains, the bending stiffness $\kappa$ was increased slowly at a rate of $8\times 10^{-6}\,\kappa / \tau$ to allow chain conformations to relax (see Section~\ref{sec:slow_dynamics}). Figure~\ref{fig:evolution_lp}(a) shows the corresponding increase in persistence lengths $L_{\text{p}} \equiv -L_\text{b}/\ln{\langle \cos{\Theta} \rangle}$, compared to the stiffness of free chains at infinite dilution, $L_\text{p,0}$. $L_\text{p}$ and $L_\text{p,0}$ increased identically up to a specific $\kappa$, but then the network strands were consistently less elongated than their free counterparts ($L_\text{p} < L_\text{p,0}$). This behavior indicates frustrated conformations (cf. Fig.~\ref{fig:increase-speed}) due to the incompatibility of the lattice spacing $a$ and persistence length $L_\text{p}$, as evidenced when plotting $L_\text{p}/a$ as a function of $\kappa$ [Fig.~\ref{fig:evolution_lp}(b)].

\begin{figure}[htb]
    \centering
    \includegraphics[width=1.00\linewidth]{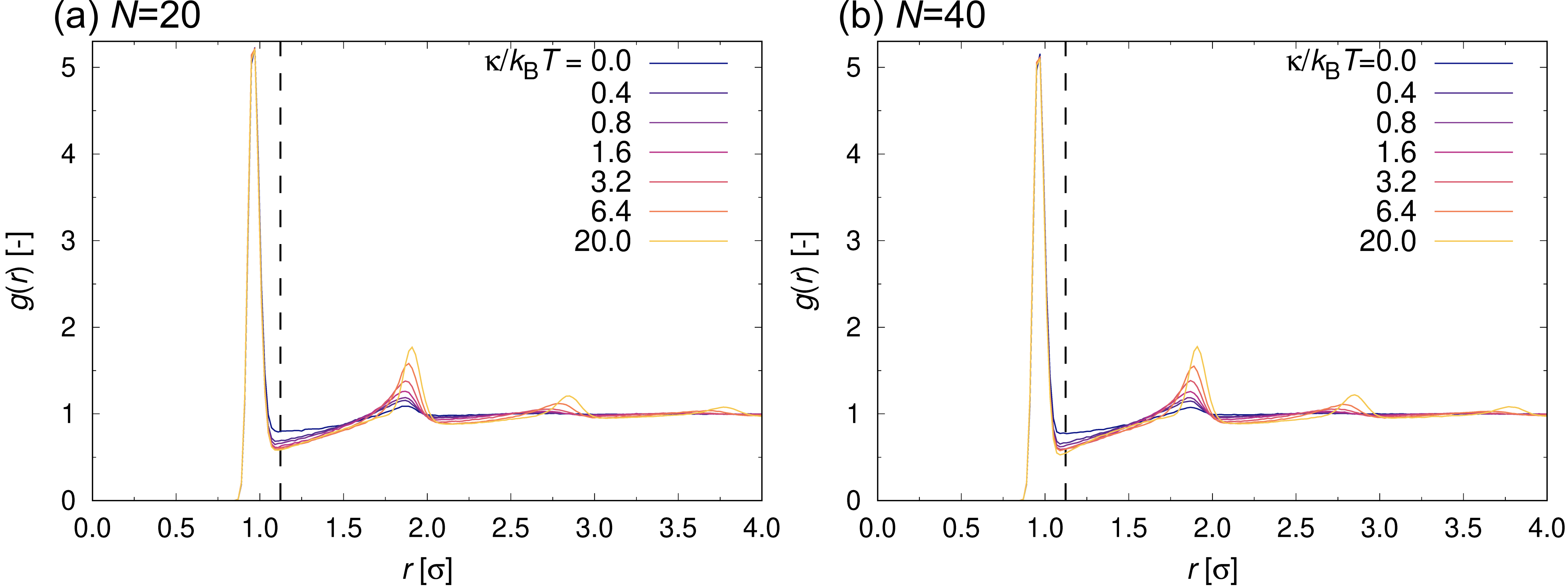}
    \caption{Radial distribution function $g(r)$ between monomers during the increase of chain bending stiffness $\kappa$ in regular networks with (a) $N=20$ and (b) $N=40$. The black dashed vertical lines describe the cutoff distance of $U_\text{WCA}$ at $r=2^{1/6}\,\sigma$.}
    \label{fig:rdf}
\end{figure}

Figure \ref{fig:rdf} shows the radial distribution functions $g(r)$ between all monomer pairs ((intra- and intermolecular)) within the regular networks as the bending rigidity $\kappa$ was gradually increased.
The probability of finding monomers within the the cutoff distance of the excluded volume interactions $U_\text{WCA}(r)$ ($r= 2^{1/6}\,\sigma$) decreases progressively as the chains become stiffer and adopt more elongated conformations.
At the same time, $g(r)$ shows more pronounced secondary peaks at intervals of $L_\text{b} \approx 0.97\,\sigma$, corresponding to the second, third, etc. neighbors along the chains.

\clearpage
\section{Hysteresis of Polymer Networks with Semi-flexible Chains}
\begin{figure}[htb]
    \centering
    \includegraphics[width=0.60\linewidth]{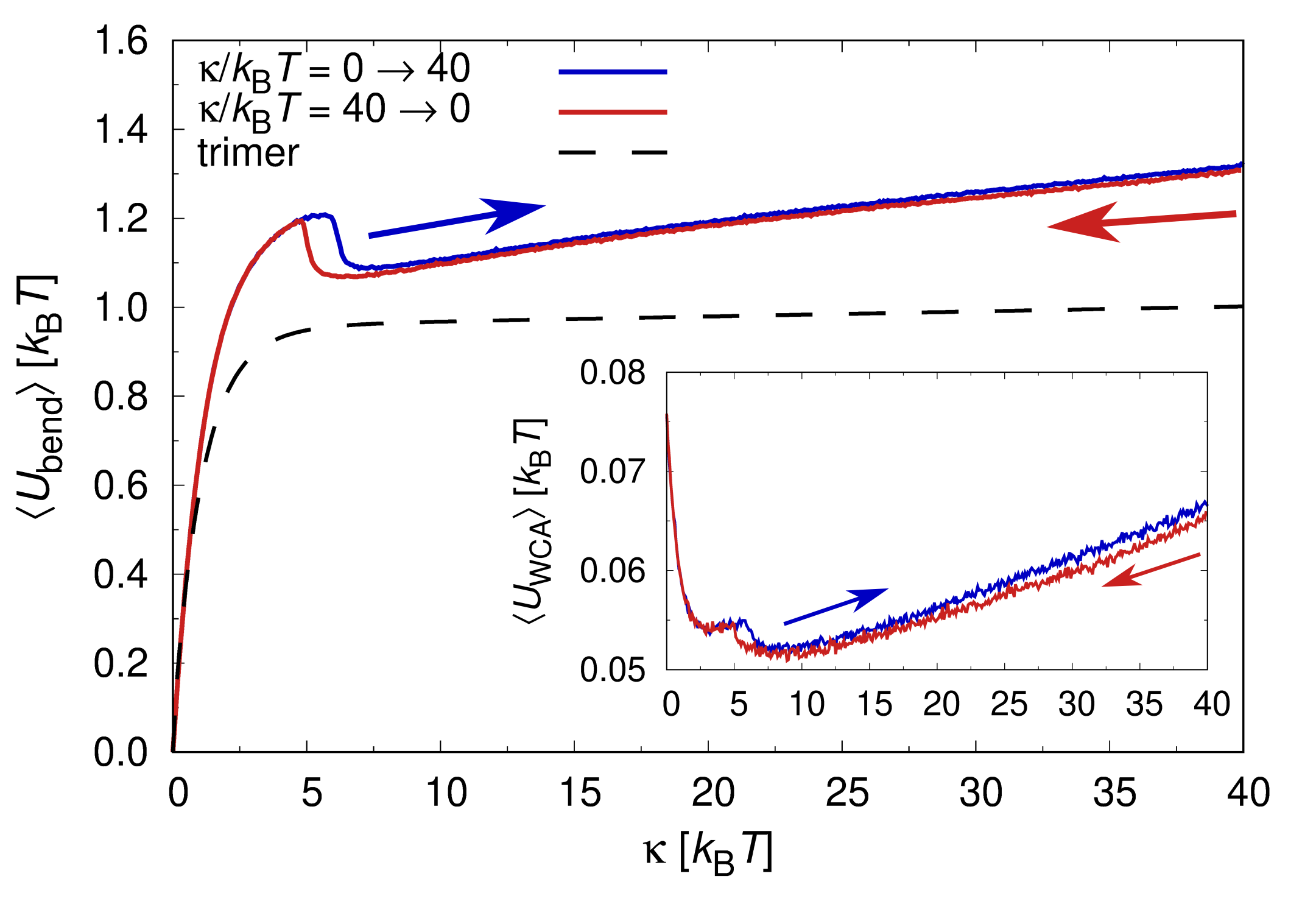}
    \caption{Evolution of average bending potential energy per triplet, $\la U_\text{bend} \ra$, for increasing (blue) and decreasing (red) bending stiffness $\kappa$ in a good solvent. Inset: Evolution of average excluded volume interaction energy per monomer, $\la U_\text{WCA} \ra$. The bending stiffness $\kappa$ was gradually changed at a rate of $8 \times 10^{-6} \kappa/\tau$ in the regular networks with $N=20$.}
    \label{fig:hysteresis_good}
\end{figure}

Figure \ref{fig:hysteresis_good} shows the evolution of the potential energies for systems where the bending stiffness $\kappa$ was increased from $\kappa = 0$ to $\kappa = 40\,k_\text{B}T$ and vice versa under good solvent conditions.
Both the bending energy per triplet, $\la U_\text{bend} \ra$, and excluded volume interaction energies, $\la U_\text{WCA} \ra$, exhibit slight hysteresis, suggesting that the relaxation of chain conformations is path-dependent and not fully reversible within the simulation timescale.

\begin{figure*}[htb]
    \centering
    \includegraphics[width=1.00\linewidth]{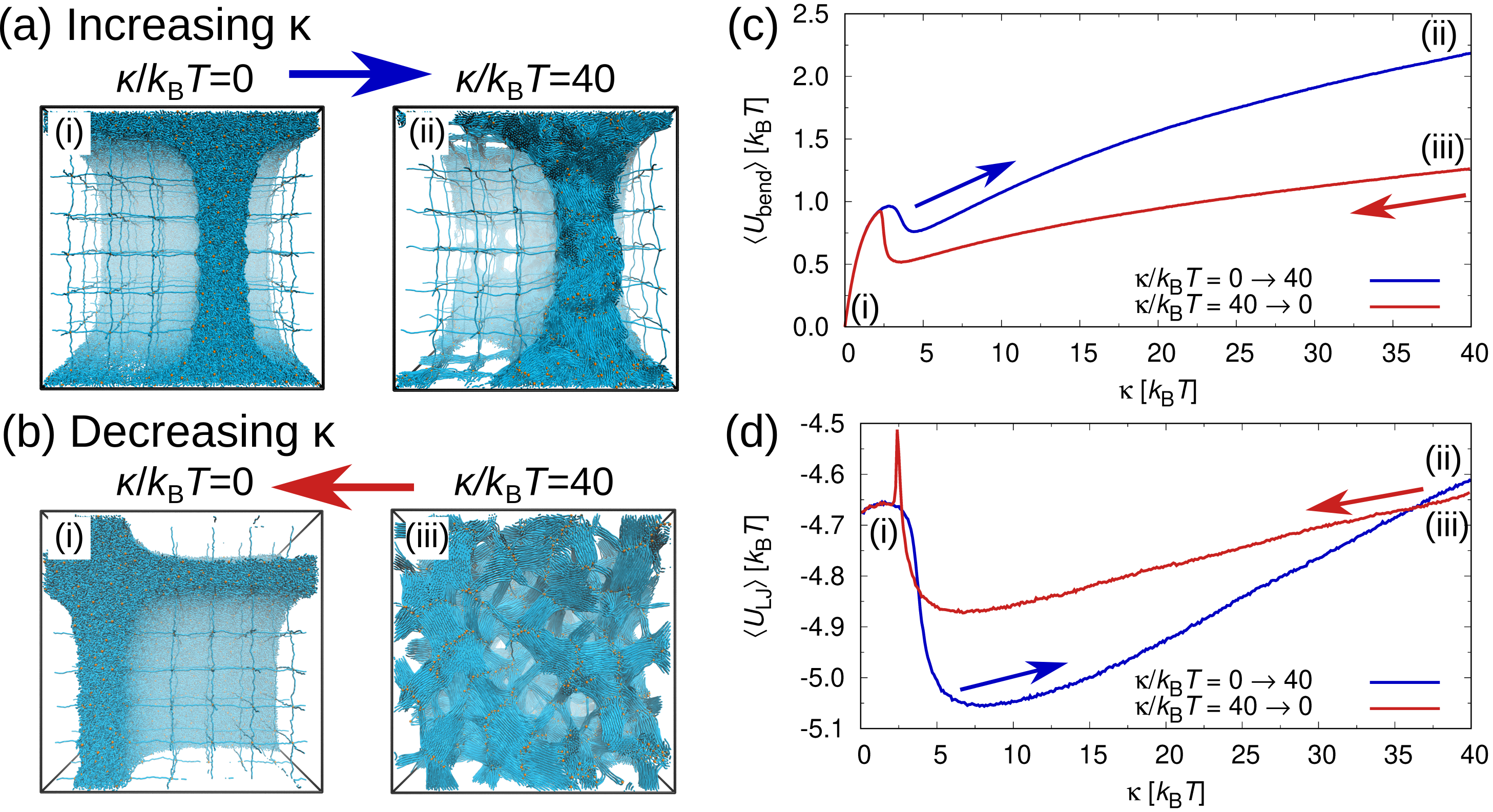}
    \caption{(a) Representative snapshots of regular networks in poor solvents when increasing bending stiffness from $\kappa=0$ to $\kappa=40\,k_\text{B}T$. (b) Same as (a) but for decreasing stiffness from $\kappa=40\,k_\text{B}T$ to $\kappa =0$. (c) Evolution of average bending potential energy per triplet, $\la U_\text{bend} \ra$ with increasing (blue) and decreasing (red) bending stiffness $\kappa$. (d) Same as (c) but for monomer-monomer interaction energy per monomer $\la U_\text{LJ} \ra$. The bending stiffness $\kappa$ was gradually changed at a rate $8 \times 10^{-6}\,\kappa/\tau$ in regular networks with $N=20$.}
    \label{fig:hysteresis_poor}
\end{figure*}

Figure \ref{fig:hysteresis_poor} shows the corresponding behavior for a network in a poor solvent. When the networks are initialized with fully flexible strands ($\kappa = 0$) in a macrophase separated state, they remain in this state with increasing bending stiffness $\kappa$ [Fig.~\ref{fig:hysteresis_poor}(a)]. However, when starting with stiff strands in a microphase separated state, the network undergoes macrophase separation as $\kappa$ is decreased  [Fig.~\ref{fig:hysteresis_poor}(b)]. For networks where $\kappa$ was increased under poor solvent conditions, the average bending potential energy $\la U_{\text{bend}} \ra$ was much higher compared to systems where the stiffness was increased in a good solvent and then changed to a poor solvent [Fig.~\ref{fig:hysteresis_poor}(c)]. This discrepancy indicates that the polymer chains adopted unfavorable conformations, which were arrested by the strong monomer–monomer attractions [Fig.~\ref{fig:hysteresis_poor}(d)]. These observations suggest that when monomer–monomer attractions are dominant, the resulting polymer network morphologies depend strongly on the system’s prior structural state.

\clearpage
\section{Nematic Ordering}
\subsection{Global Nematic Ordering in Polymer Solutions}
\begin{figure}[htb]
    \centering
    \includegraphics[width=0.65\linewidth]{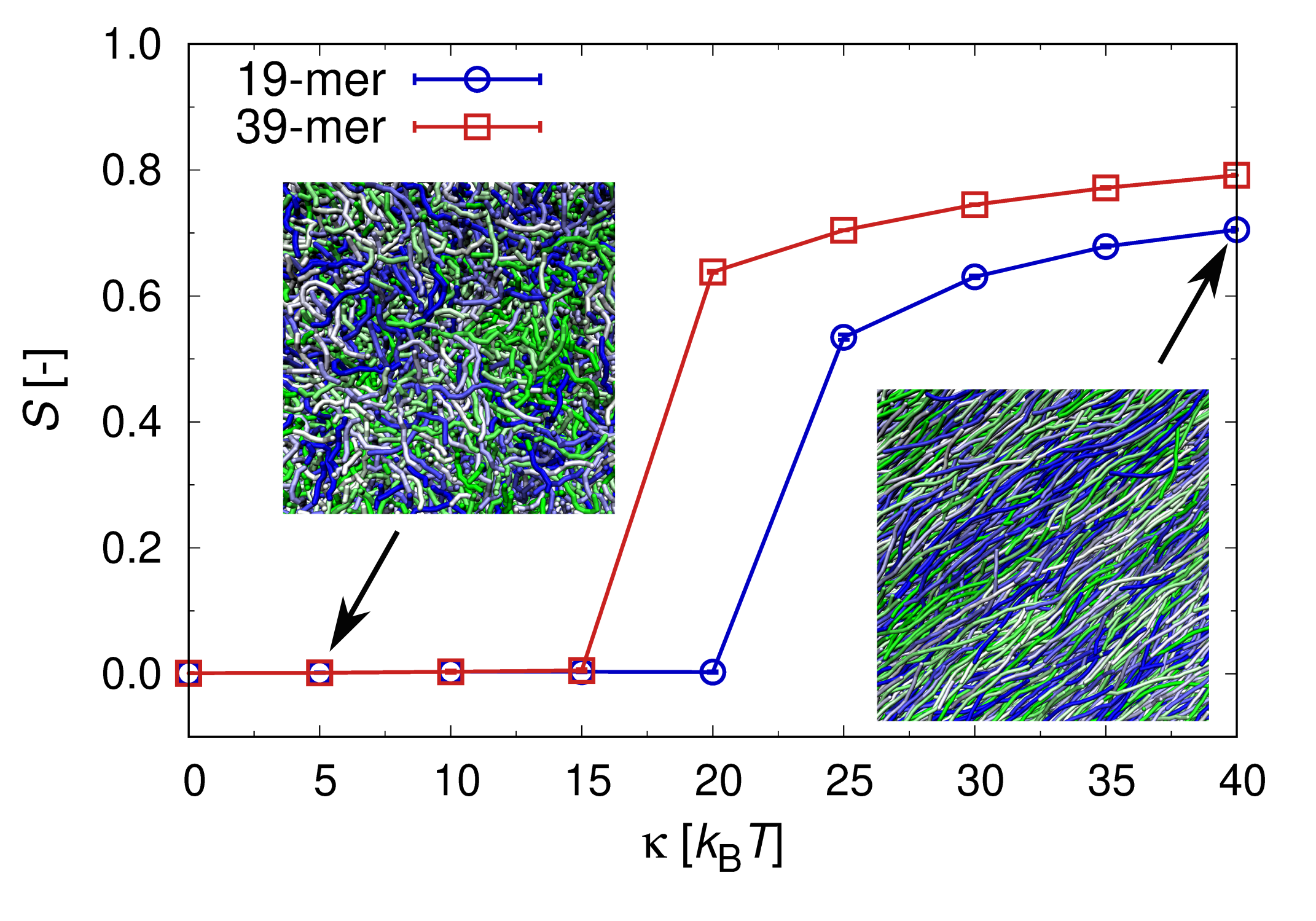}
    \caption{Global nematic order parameter $S$ vs. chain bending stiffness $\kappa$ in simple homopolymer solutions at $\rho=0.4\,\sigma^{-3}$.}
    \label{fig:global_nematic}
\end{figure}

To characterize the conformational behavior of semi-flexible homopolymers in good solvents, we examined a range of chain bending rigidities and chain lengths. For the comparison with the polymer networks, we simulated 19-mers and 39-mers at the same monomer number density $\rho=0.4\,\sigma^{-3}$ and temperature $k_\text{B}T=\varepsilon$. The size of the cubic simulation boxes were set to a volume $V=(80.00\,\sigma)^{3}$ for computational efficiency. We computed the global nematic order parameter $S$ from the orientation tensor
\begin{equation}
Q^{\alpha \beta} = \frac{1}{2} \left\langle  3 \mathbf{u}^{\alpha}_{ij} \mathbf{u}^{\beta}_{ij} - \delta^{\alpha \beta} \right\rangle,
    \label{eq:orientation_tensor}
\end{equation}
where $\alpha$ and $\beta$ represent the Cartesian components $x$, $y$, $z$, and $\mathbf{u}_{ij}$ denotes the unit bond vector between monomers $i$ and $j$. To characterize the global nematic order parameter, the average was taken both temporal and over all the bonds in the system. The tensor $Q^{\alpha \beta}$ has three eigenvalues $\lambda_{1} \geq \lambda_{2} \geq \lambda_{3}$, and we defined the global nematic order parameter as $S \equiv \lambda_{1}$. Figure \ref{fig:global_nematic} shows the relationship between $S$ and the chain bending stiffness $\kappa$, starting from a perfectly ordered configuration. Global nematic ordering was observed above $\kappa/k_{\text{B}}T = 25$ for 19-mer and above $\kappa/k_{\text{B}}T = 20$ for 39-mer.

\subsection{Local Nematic Ordering in Polymer Networks}
\begin{figure*}[htb]
    \centering
    \includegraphics[width=1.00\linewidth]{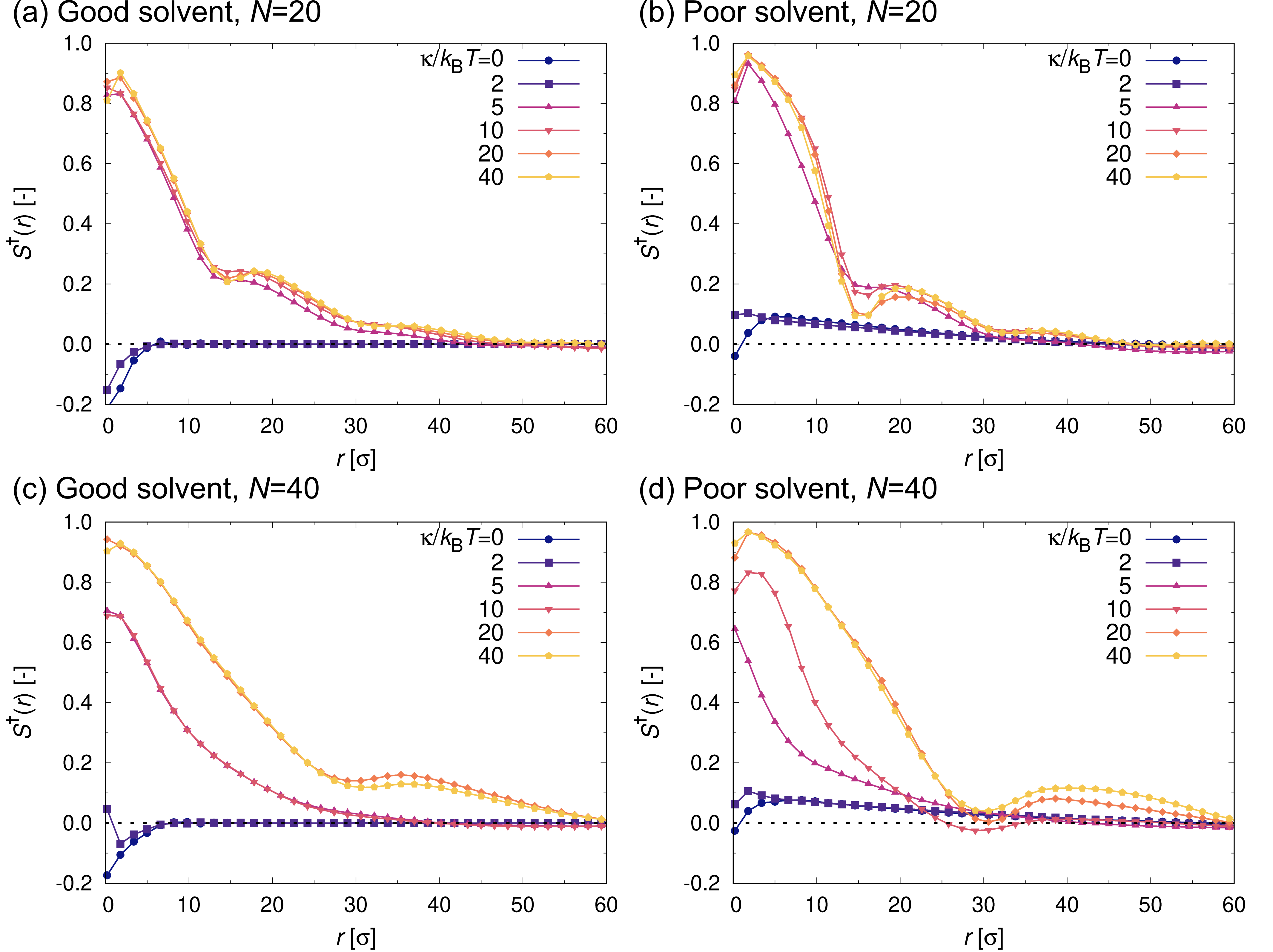}
    \caption{Local nematic order parameter $S^\dagger$ vs. distance $r$ between the centers of mass of two individual chains in regular networks at varying chain bending stiffness $\kappa$. Panels show data for chains with $N=20$ in (a) good and (b) poor solvent, and for $N=40$ in (c) good and (d) poor solvents.}
    \label{fig:local_nematic}
\end{figure*}

Having established the global nematic ordering in polymer solutions, we next analyzed the local nematic ordering $S^\dagger$ of semi-flexible polymers in the regular networks:
\begin{equation}
S^{\dagger} \left( r \right) = \frac{1}{2} \langle 3 \cos^{2}{\theta_{ij}} - 1 \rangle ,
    \label{eq:S_local}
\end{equation}
where $\theta_{ij}$ is the angle between the end-to-end vectors of two individual chains $i$ and $j$, whose center of masses are separated by a distance $r = |\mathbf{r}_{j} - \mathbf{r}_{i}|$. Figure \ref{fig:local_nematic} shows $S^\dagger$ for various solvent qualities, chain lengths $N$, and chain bending stiffnesses $\kappa$.

\clearpage
\section{Chain Conformations in Good Solvent Conditions}
\begin{figure}[htb]
    \centering
    \includegraphics[width=0.60\linewidth]{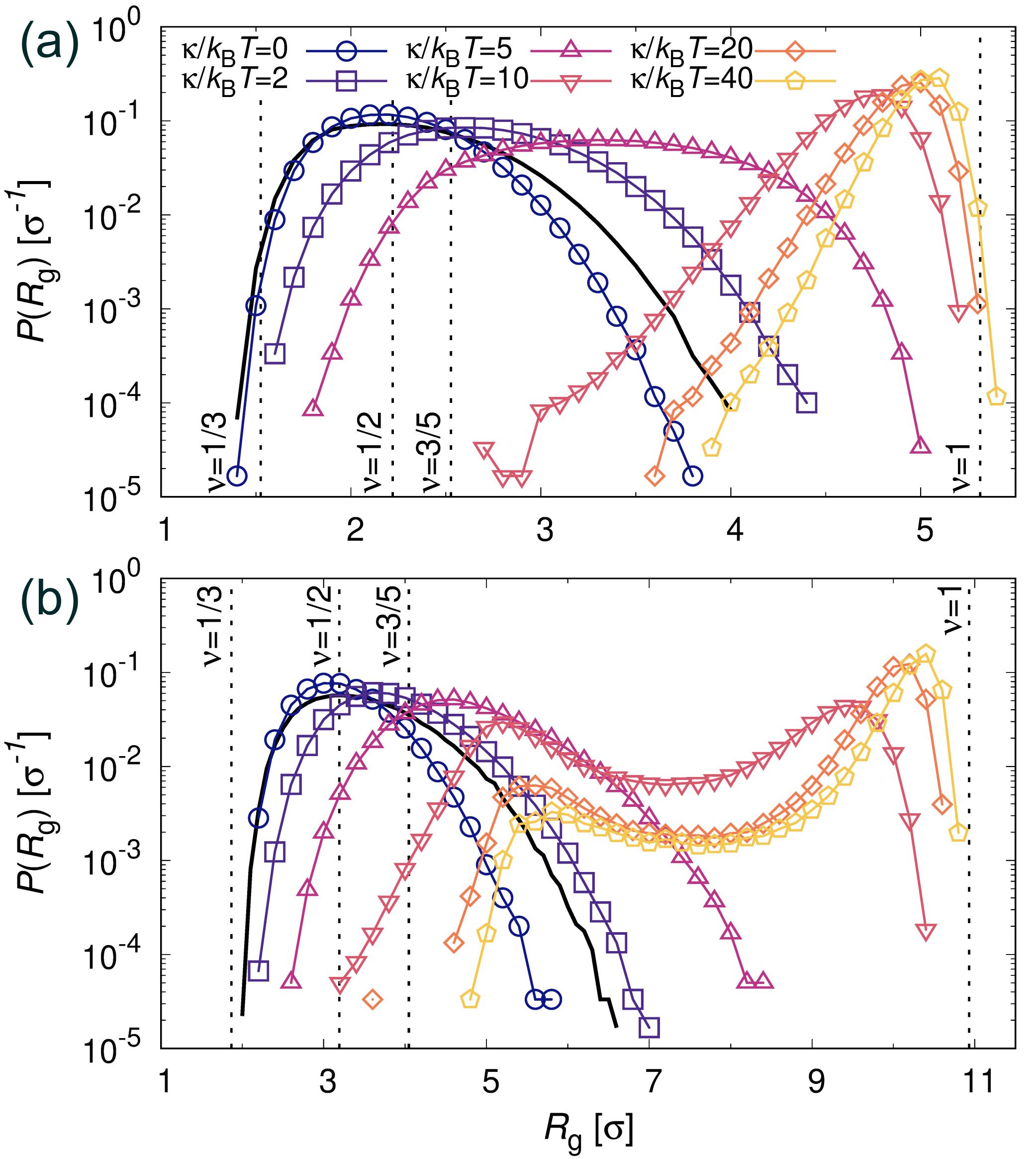}
    \caption{Probability distribution of the radius of gyrations $P(R_{\text{g}})$ of chains in a good solvent for various bending stiffnesses $\kappa$, with (a) chain length $N=20$, and (b) $N=40$. The black curves show $P(R_{\text{g}})$ of simple homopolymer solutions of (a) 19-mer and (b) 39-mer at the same monomer number density $\rho = 0.4\,\sigma^{-3}$ in good solvents. Black dotted vertical lines indicate the computed $R_{\text{g}}$ values corresponding to different characteristic chain conformations characterized by their Flory's exponents $\nu$.} 
    \label{fig:Rg_goodSolvent}
\end{figure}

Figure \ref{fig:Rg_goodSolvent} shows the probability distribution of the radius of gyration $P(R_{\text{g}})$ for regular networks in a good solvent. For $\kappa=0$, the network strands have similar conformations as fully flexible polymers in solution, but with a slightly smaller $R_\text{g}$ due to the constraints from cross-links. With increasing stiffness $\kappa$, the population of coil-like chains diminishes drastically, and more chains tend to adopt expanded conformations. At $\kappa = 5\,k_\text{B}T$, a small fraction of chains with coil-like conformations remains, but stretched chains dominate. For $\kappa \geq 10\,k_\text{B}T$, $P(R_\text{g})$ shifts to the fully stretched state and becomes more narrow. For longer semi-flexible chains ($N=40$, $L_\text{c}/a = 5.83$), a fraction of chains is bent due to the mismatches in $L_\text{c}$, $L_\text{p}$ and $a$, resulting in the additional peak of $P(R_\text{g})$ at intermediate $R_\text{g}$ [Fig. \ref{fig:Rg_goodSolvent}(b)].

\begin{figure}[htb]
    \centering
    \includegraphics[width=0.60\linewidth]{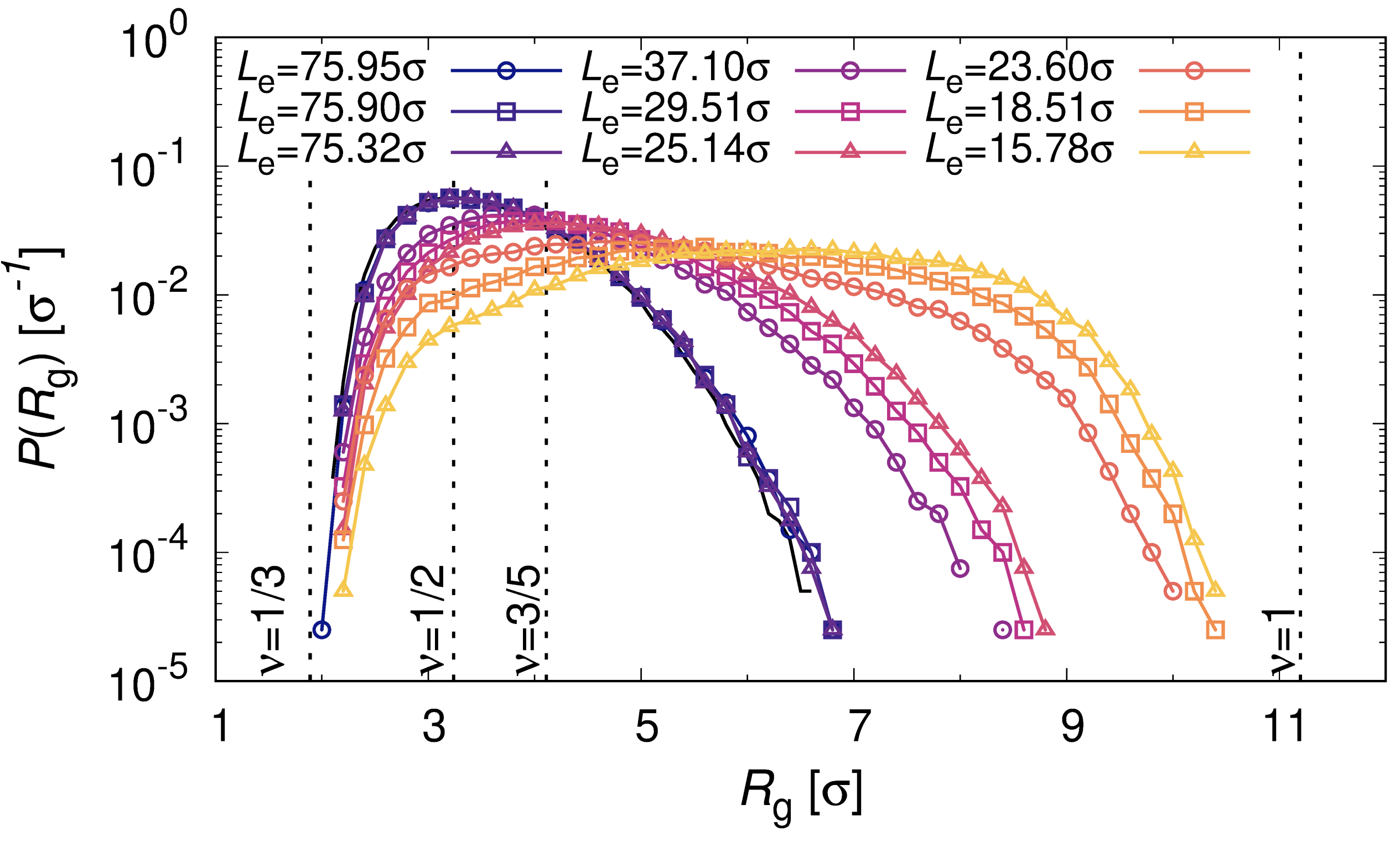}
    \caption{Probability distribution of the radius of gyrations $P(R_{\text{g}})$ of chains in entangled networks with different entanglement lengths $L_{\text{e}}$ under good solvent conditions. The circle, square, triangle points correspond to cross-linking fraction $\phi_\text{c}=0.0690,\ 0.0862, \ 0.1034$, respectively. The black curves show $P(R_{\text{g}})$ of simple homopolymer solutions of $N=40$ at the same monomer number density $\rho = 0.4\,\sigma^{-3}$ in good solvents. Black dotted vertical lines indicate the computed $R_{\text{g}}$ values corresponding to different characteristic chain conformations characterized by their Flory's exponents $\nu$.}
    \label{fig:Rg_Ne_good}
\end{figure}

Figure \ref{fig:Rg_Ne_good} presents $P(R_{\text{g}})$ for entangled networks in good solvents with different entanglement length $L_{\text{e}}$. For weakly entangled networks ($L_{\text{e}} \approx 75\,\sigma$), the chain conformations match those in homopolymer solutions with the same chain length ($N=40$).  With decreasing $L_{\text{e}}$, the distribution $P(R_{\text{g}})$ becomes broader, reflecting an increased occurrence of highly expanded conformations.

\clearpage
\section{Solvent quality dependence on the phase separation behaviors}
\begin{figure}[htb]
    \centering
    \includegraphics[width=1.00\linewidth]{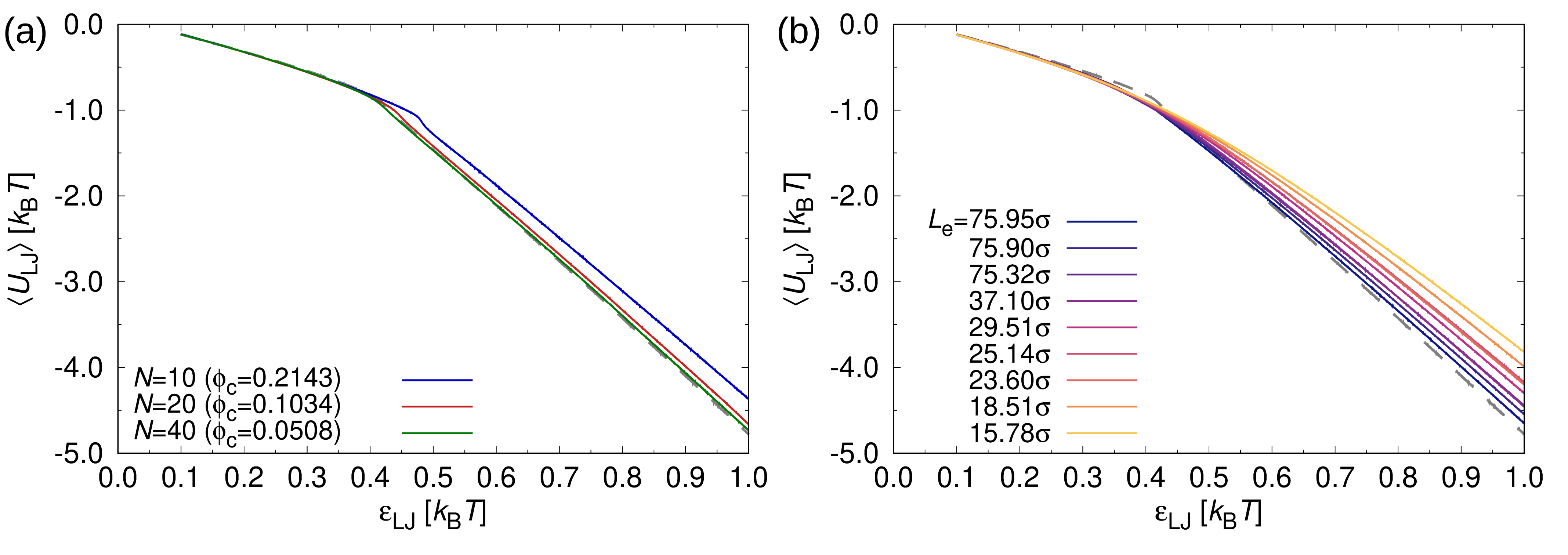}
    \caption{Evolution of the average Lennard-Jones potential energy per monomer $\la U_\text{LJ} \ra$. The interaction strength $\varepsilon_{\text{LJ}}$ was gradually increased at a rate of $2 \times 10^{-6}\,\varepsilon_{\text{LJ}}/\tau$ in both (a) regular and (b) entangled networks. Black dotted curves correspond to the $\la U_{\text{LJ}} \ra$ vs $\varepsilon_{\text{LJ}}$ for simple flexible hompolymer solutions (40-mers) at the same monomer number density $\rho=0.4\,\sigma^{-3}$.}
    \label{fig:LJpot}
\end{figure}

\begin{figure}[htb]
    \centering
    \includegraphics[width=1.00\linewidth]{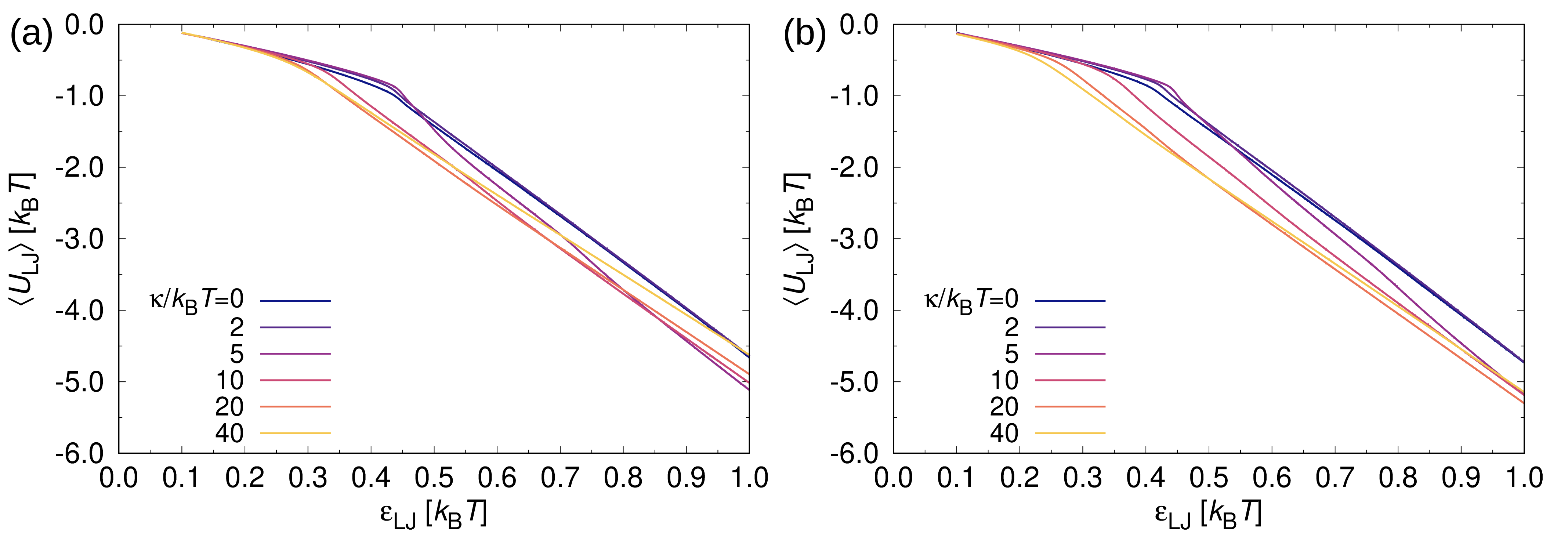}
    \caption{Evolution of the average Lennard-Jones potential energy per monomer $\la U_\text{LJ} \ra$ for regular networks with semi-flexible chains. The interaction strength $\varepsilon_{\text{LJ}}$ was gradually increased at a rate of $2 \times 10^{-6}\,\varepsilon_{\text{LJ}}/\tau$ for both (a) $N=20$ and (b) $N=40$.}
    \label{fig:LJpot_rigidity}
\end{figure}

Phase separation was triggered by progressively increasing the attractive Lennard–Jones interaction strength $\varepsilon_\text{LJ}$, which effectively reduces the solvent quality. For polymer networks with flexible chains ($\kappa =0$), the transition was observed for $0.4\,k_\text{B}T \lesssim \varepsilon_\text{LJ} \lesssim 0.5\,k_\text{B}T$ in the regular networks, evidenced by a clear change in slope in the decrease of $\la U_{\text{LJ}} \ra$ [Fig. \ref{fig:LJpot}(a)]. The transition weakly depends on the network state -- specifically the cross-linking fraction and degree of entanglements -- such that networks with stronger topological constraints (higher $\phi_{\text{c}}$ or shorter $L_{\text{e}}$) phase separated at slightly higher $\varepsilon_{\text{LJ}}$ [Fig. \ref{fig:LJpot}(b)]. Regular networks with semi-flexible strands ($\kappa \geq 10\,k_\text{B}T$) phase separate at lower $\varepsilon_{\text{LJ}}$ values, as the chains already exhibit local nematic ordering in good solvents, which facilitates the formation of nematic bundles (Fig. \ref{fig:LJpot_rigidity}). 

\clearpage
\section{Pore Size Distribution Calculation}
\begin{figure}[htb]
    \centering
    \includegraphics[width=0.60\linewidth]{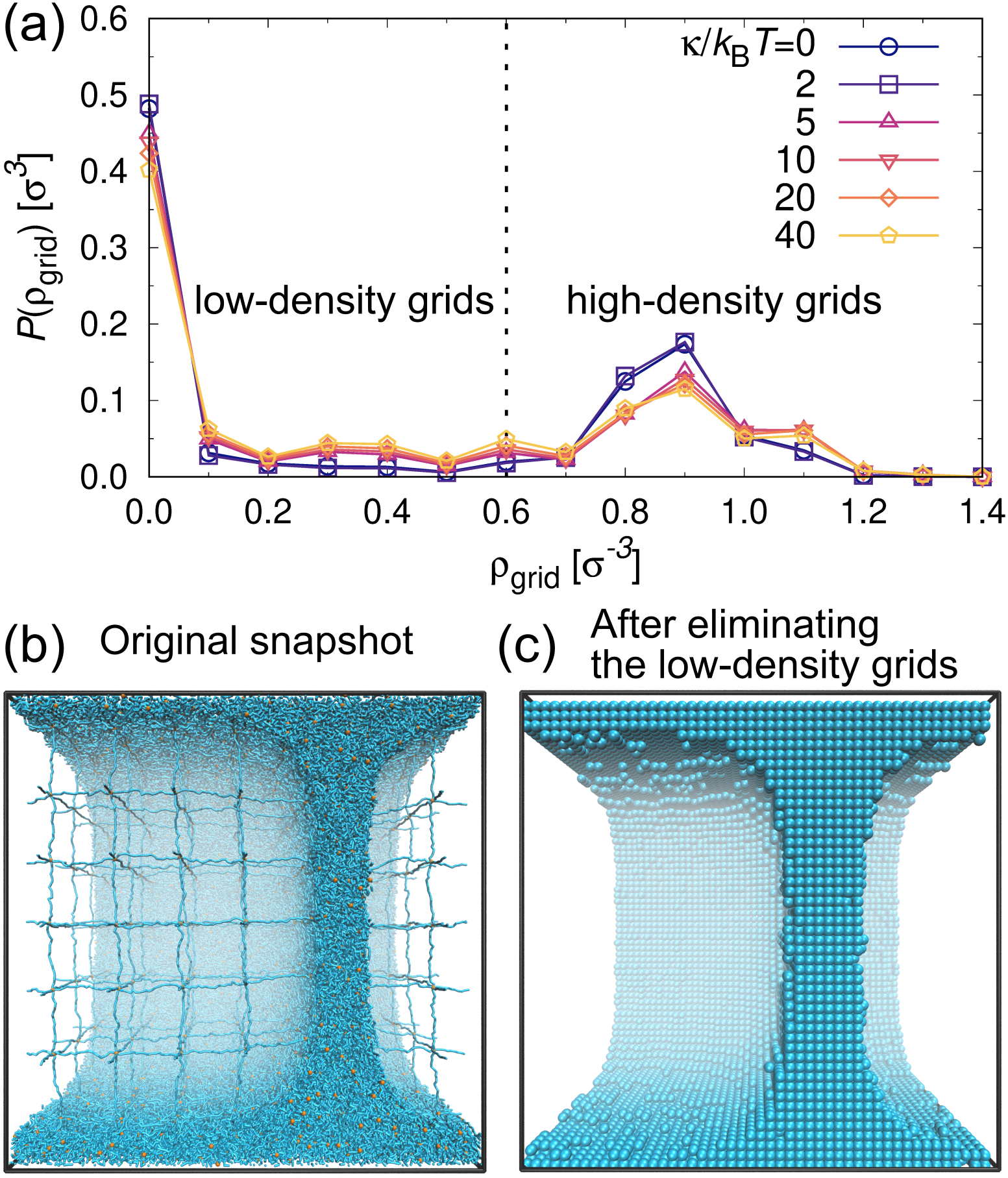}
    \caption{(a) Probability distribution of local density $P(\rho_{\text{grid}})$ of the phase-separated systems with chain length $N=20$ and varying bending stiffness $\kappa$. The black dotted line at $\rho_{\text{grid}}=0.6\,\sigma^{-3}$ indicates the threshold for dividing cells into empty and full cells, respectively. (b,c) Representative snapshots of a phase-separated system (b) before and (c) after coarse-graining into cubic cells.}
    \label{fig:method_psd}
\end{figure}

To characterize the size of the low-density regions, we applied the geometric pore size distribution method defined by Gelb and Gubbins.\cite{gelb:lng:1999, bhattacharya:lng:2006} To this end, we first create a coarse representation of the system, where the system is divided into cubic cells with an edge length of $2.5\,\sigma$. We then calculated the local monomer number density $\rho_{\text{grid}}$ inside each cube [Fig.~\ref{fig:method_psd}(a)], and identified cells with $\rho_{\text{grid}} \leq 0.6\,\sigma^{-3}$ as empty and as full otherwise [Fig.~\ref{fig:method_psd}(b,c)]. Then, we placed $50,000$ spherical probes into this coarse representation, and optimized their position and diameter $d_\text{pore}$ such that they still encompasses the original insertion point $\mathbf{r}$ without overlapping the surrounding filled cells.

\begin{figure}[htb]
    \centering
    \includegraphics[width=1.00\linewidth]{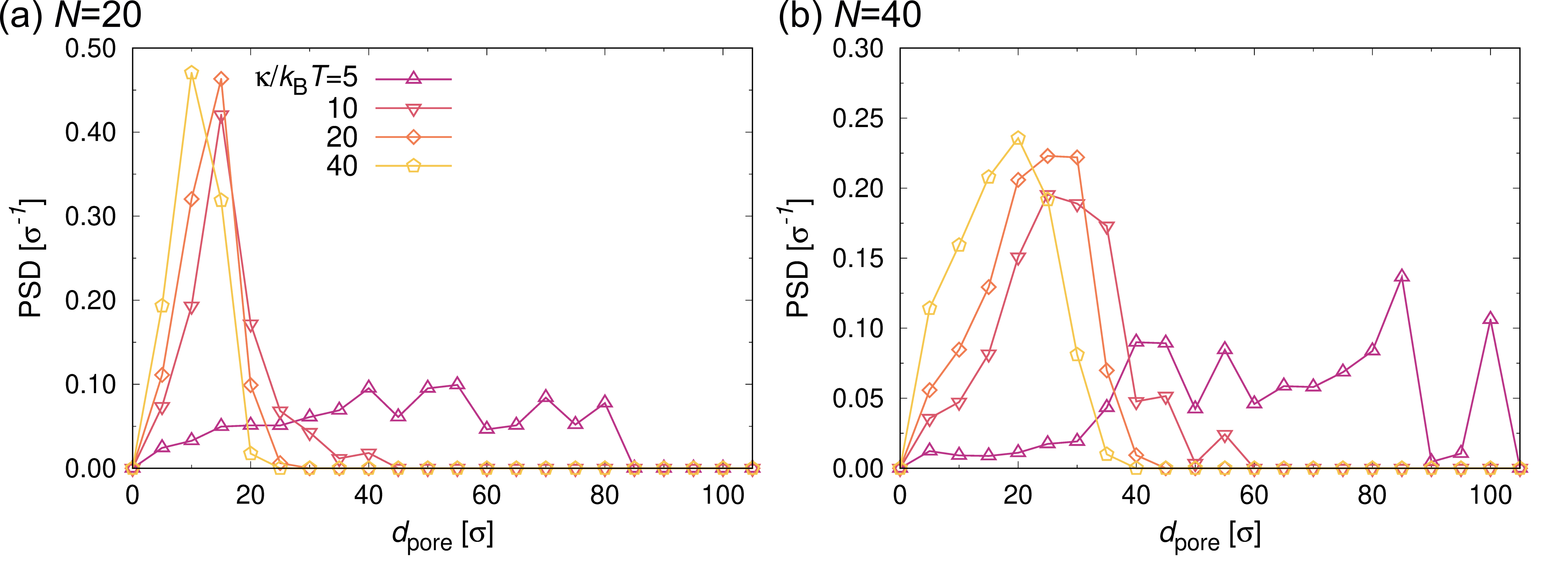}
    \caption{Pore size distribution (PSD) in regular networks with semi-flexible chains under poor solvent conditions for various bending stiffnesses $\kappa$ with (a) chain length $N=20$, and (b) $N=40$.}
    \label{fig:psd_semi-flexible}
\end{figure}

\begin{figure}[htb]
    \centering
    \includegraphics[width=0.55\linewidth]{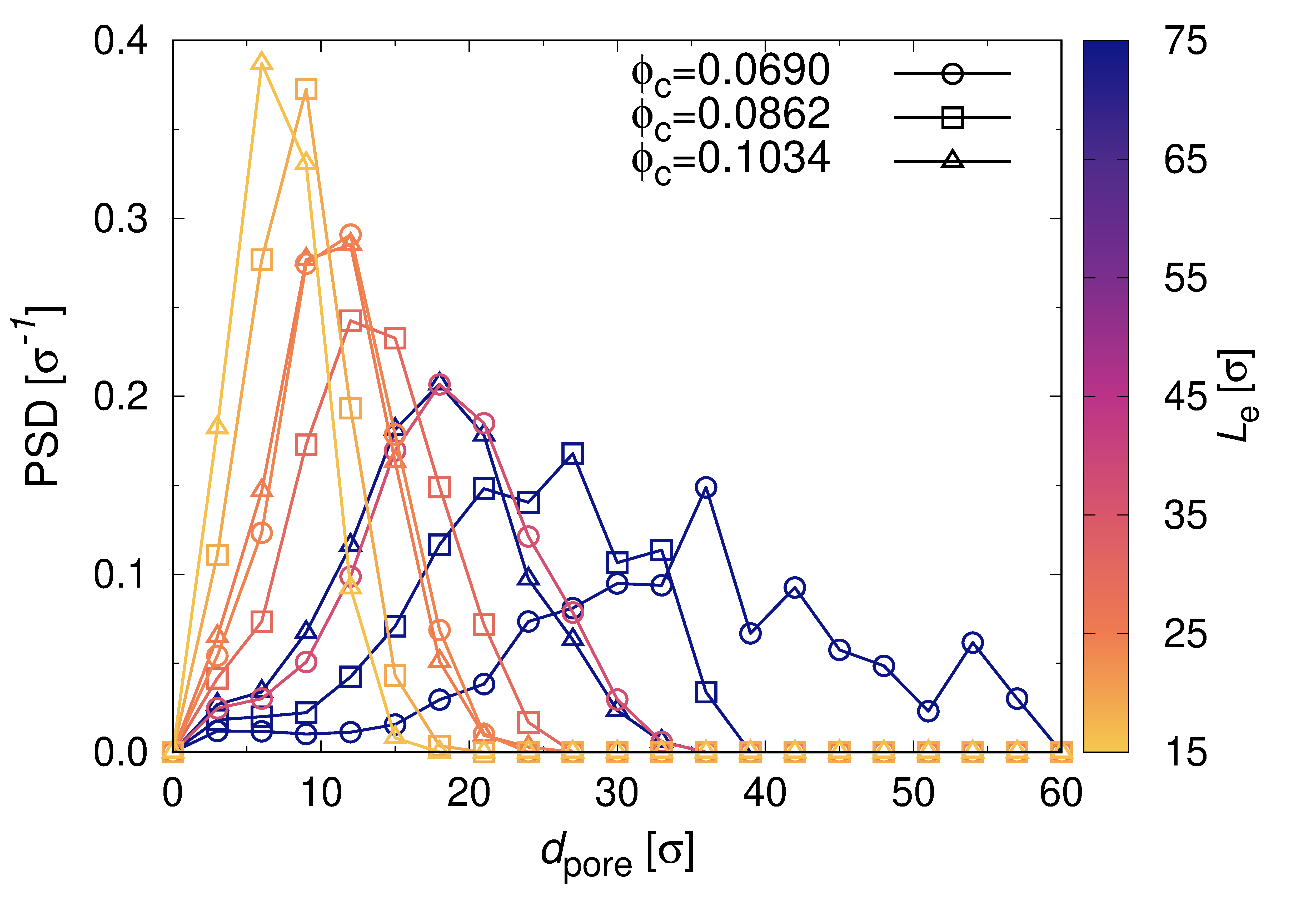}
    \caption{Pore size distribution (PSD) in entangled networks with flexible chains under poor solvent conditions for various entanglement lengths $L_{\text{e}}$. }
    \label{fig:psd_entangle}
\end{figure}

The phase separation length scale was then defined as the average of pore diameters $\la d_\text{pore} \ra$. Since the individual probes are from independent trials, they can overlap, causing an elongated pore to appear as several overlapping probes in sequence. Figure~\ref{fig:psd_semi-flexible} shows the pore size distribution (PSD) for the regular networks with different bending stiffnesses $\kappa$ in a poor solvent. As $\kappa$ increases, the PSD shifts towards smaller $d_\text{pore}$. Figure \ref{fig:psd_entangle} shows the PSD of entangled networks with different entanglement lengths $L_{\text{e}}$ in a poor solvent. As $L_\text{e}$ decreases, the PSD shifts toward smaller $d_{\text{pore}}$, indicating that more highly entangled networks exhibit smaller pores.

\section{Polydisperse regular lattice exhibit macrophase separation}
\begin{figure}[htb]
    \centering
    \includegraphics[width=1.00\linewidth]{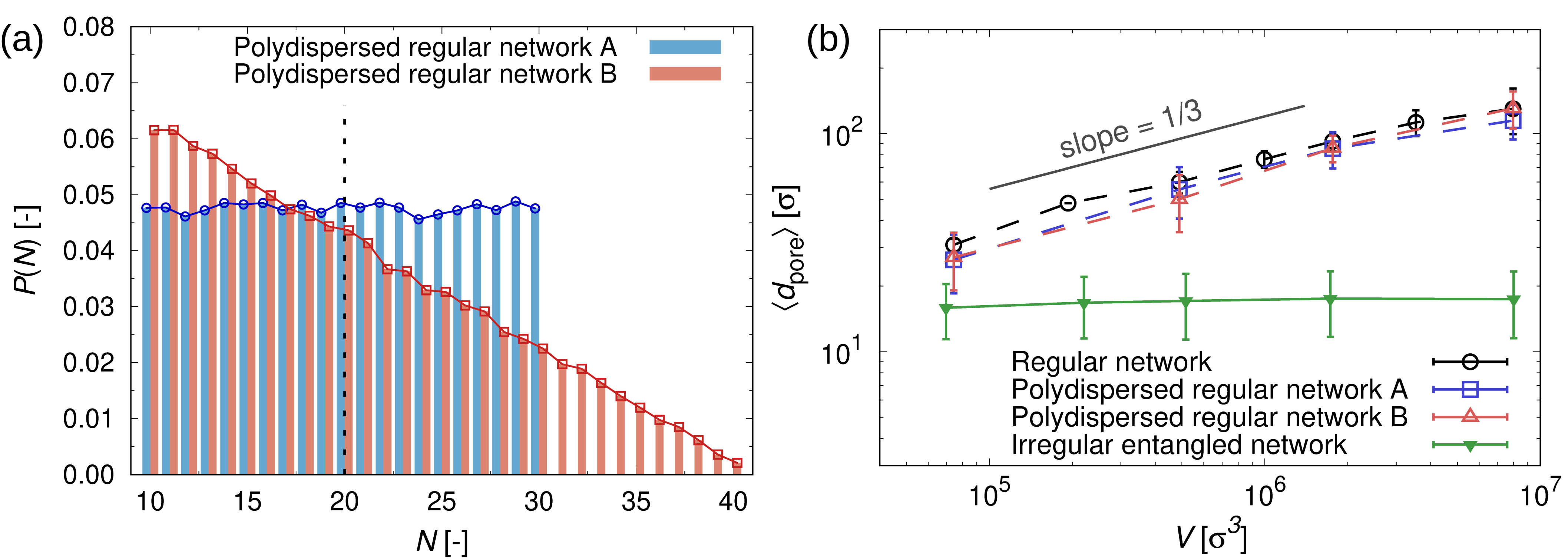}
    \caption{(a) Probability distribution of chain length $P(N)$ within regular networks with different polydispersity A and B, at the same number-averaged chain length $\langle N \rangle=20$. (b) Average pore size $\langle d_{\text{pore}} \rangle$ as function of the system volume $V$ for regular networks of (polydisperse) flexible chains (dashed lines and open symbols) and entangled networks of monodisperse flexible chains (solid lines and filled symbols). All data shown for the same cross-linking fraction $\phi_{\text{c}}=0.1034$.}
    \label{fig:polydispersed-Lattice}
\end{figure}

\clearpage
\section{Contributions to the overall bulk modulus}
\begin{figure}[htb]
    \centering
    \includegraphics[width=1.00\linewidth]{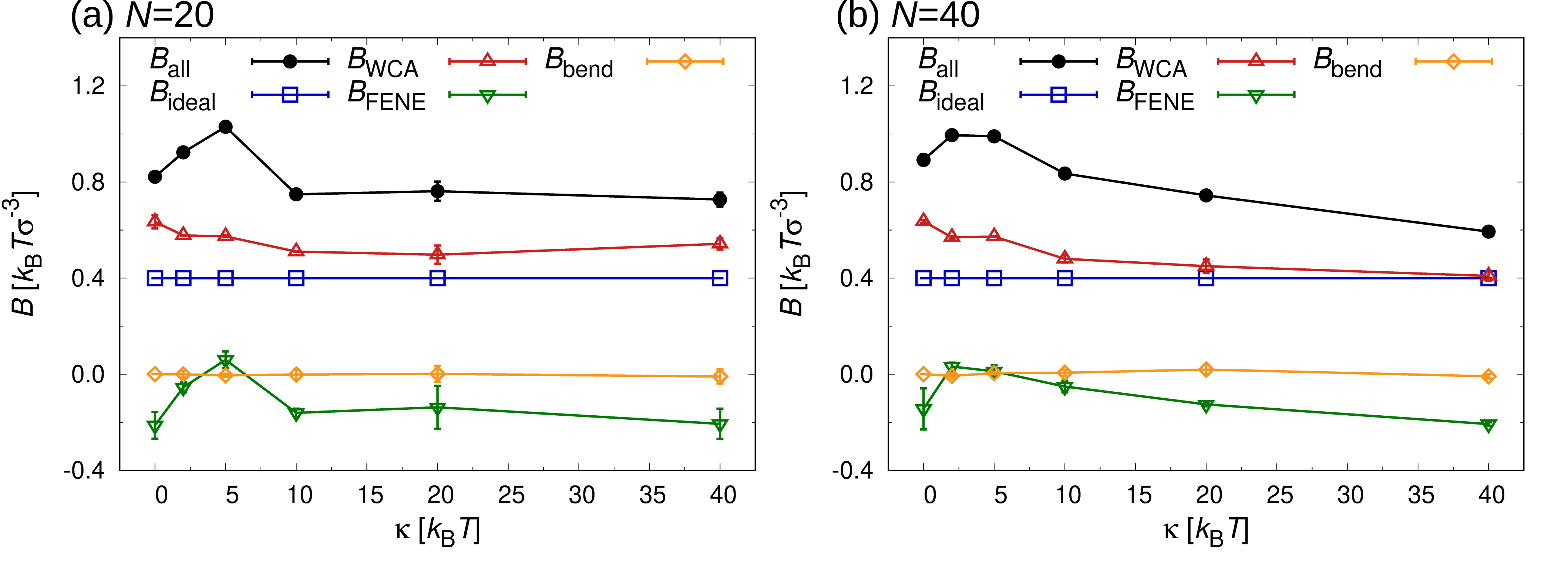}
    \caption{Bulk modulus $B_\text{all}$, and its contributions from from the ideal gas term $B_{\text{ideal}}$, excluded-volume interactions $B_{\text{WCA}}$, bonding potential $B_{\text{FENE}}$, and bending potential $B_{\text{bend}}$ as functions of the chain bending stiffness parameter $\kappa$ for regular networks with semi-flexible chains.}
    \label{fig:Bindiv-rigidity}
\end{figure}

\begin{figure}[htb]
    \centering
    \includegraphics[width=0.55\linewidth]{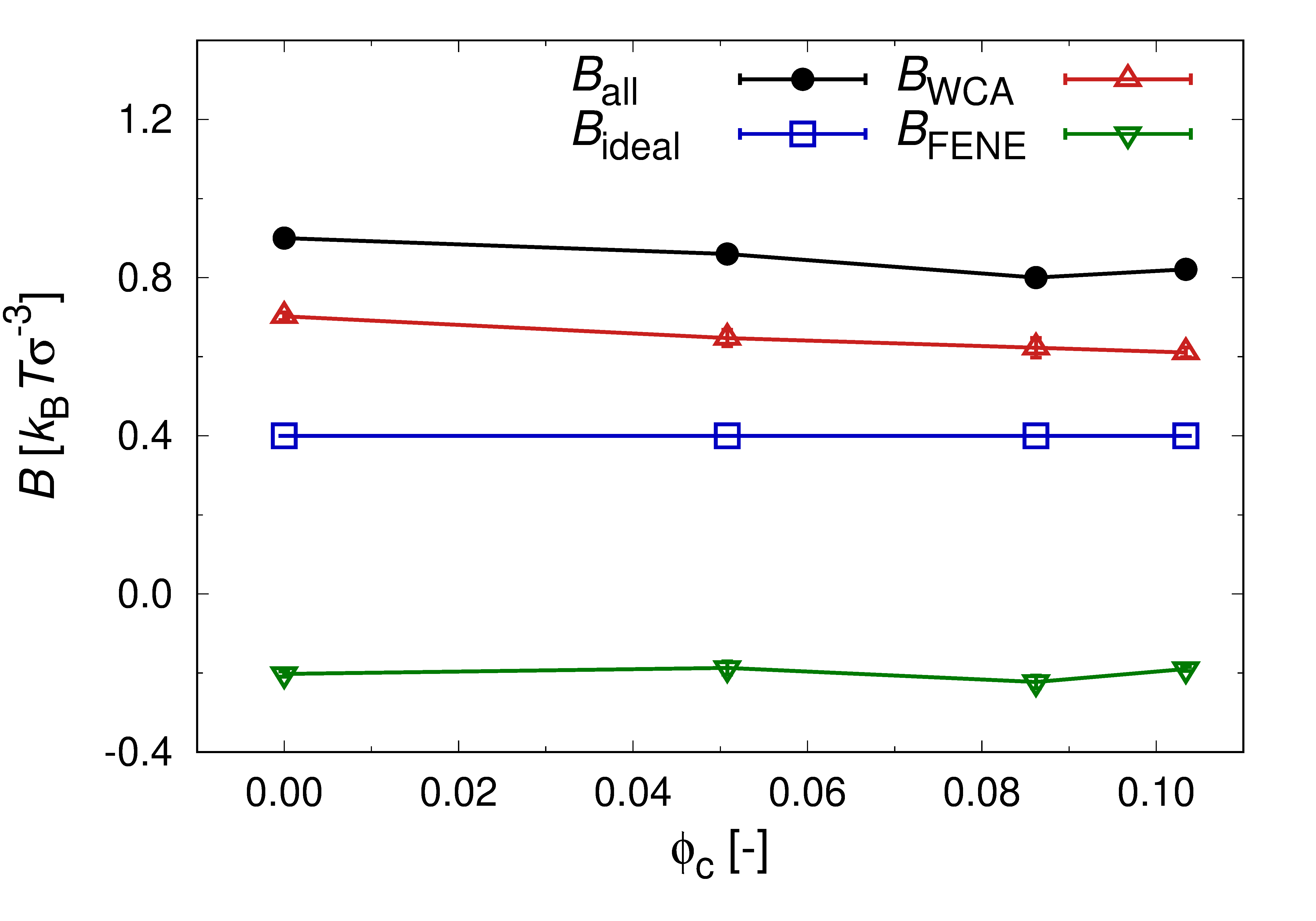}
    \caption{Bulk modulus $B_\text{all}$, and its contributions from from the ideal gas term $B_{\text{ideal}}$, excluded-volume interactions $B_{\text{WCA}}$, and bonding potential $B_{\text{FENE}}$ as functions of cross-linking fraction $\phi_{\text{c}}$ for entangled networks. Points at $\phi_{\text{c}}=0$ correspond to the linear-chain solution with hexa-functional cross-linkers prior to cross-linking.}
    \label{fig:Bindiv-entangle}
\end{figure}

\clearpage
\section{The relationship between network elasticity and microphase separation length scale}
\begin{figure}[htb]
    \centering
    \includegraphics[width=0.55\linewidth]{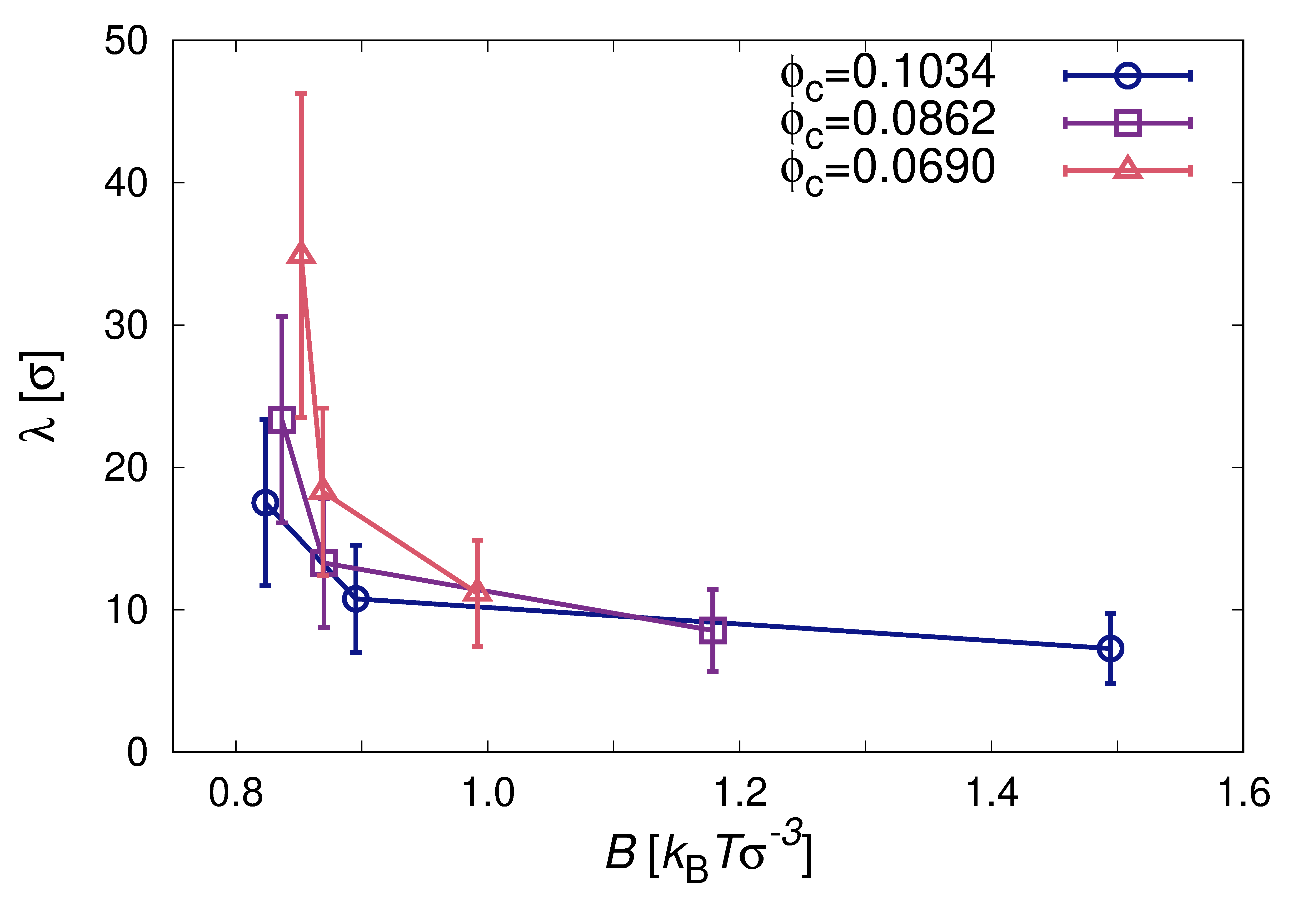}
    \caption{Characteristic length scale of microphase separation $\lambda$ as a function of bulk modulus $B$ for entangled networks with different cross-linking fraction $\phi_\text{c}$. }
    \label{fig:Lambda-B}
\end{figure}

\providecommand*{\mcitethebibliography}{\thebibliography}
\csname @ifundefined\endcsname{endmcitethebibliography}
{\let\endmcitethebibliography\endthebibliography}{}